\documentclass[aps,prc,preprint,showpacs]{revtex4}
\usepackage{amsfonts}
\usepackage{amssymb}
\usepackage{graphics}
\usepackage{graphicx}
\usepackage{color}
\usepackage{slashed}
\usepackage{enumerate}
\def\ket#1{|#1 \rangle}
\def\bra#1{\langle #1|}

\newcommand{\mc}{\multicolumn}

\begin{document}

\title{Asymmetry of the neutrino mean free path in partially spin polarized neutron matter within the Skyrme model}

\author{E. Bauer$^{1,2}$ and J. Torres Pati\~no$^{1}$}
\affiliation{$^1$ IFLP, CCT-La Plata, CONICET. Argentina}
\affiliation{$^2$Facultad de Ciencias Astron\'omicas y Geof\'{\i}sicas, Universidad Nacional de La Plata, Argentina}


\begin{abstract}
The asymmetry in the neutrino mean free path for the absorption reaction $\nu + n \rightarrow e^{-} + p$, is evaluated within hot neutron matter under a strong magnetic field. We consider densities in the range $0.05 \leq \rho \leq 0.4$ fm$^{-3}$, several temperatures up to 30 MeV and magnetic field strengths from $B=0$ up to $B=10^{18}$~G. Polarized neutron matter is described within the non--relativistic Hartree--Fock model using the LNS Skyrme interaction. The neutrino mean free path has a weak dependence with the temperature and in the strong magnetic field region, it decreases for growing values of it. This contrast with the scattering reaction $\nu + n \rightarrow \nu' + n'$, where the average mean free path is almost independent of the magnetic field and has a strong dependence with the temperature. We have evaluated the asymmetry from both the absorption and scattering reactions. Our results shows that the total asymmetry depends on the magnetic field intensity, the density and the temperature. For a density of $0.16$ fm$^{-3}$ and for a magnetic field strength of B=$10^{17}$~G, the asymmetry in the mean free path is found to be, $\sim 9\%$ and $\sim 3.4\%$ for temperatures of T$=15$ and $30$~MeV, respectively. While the same set of asymmetries for B=$10^{18}$~G, is $\sim 58\%$  and $\sim 48\%$.
\end{abstract}


\maketitle

\section{Introduction}
\label{int}
Neutrinos play an important role in the evolution of stellar objects. The physics of neutrinos is relevant at all stages of the stellar evolution, starting from the supernova explosions~\cite{bethe,janka,burrows}. After such event, the remaining matter forms a compact object where the neutrinos are one of the key elements for understanding this process~\cite{burrowsb, jankab}. There are many mechanisms which produce neutrinos in a neutron star. A complete review on this point can be found in~\cite{Ya01}. The possible reactions depends on the neutron star region under consideration. In the neutron star crust one has electron--positron annihilation ($e^{-} e^{+} \to \nu \bar{\nu}$), photon decay ($\gamma \to \nu \bar{\nu}$), electron--nucleus bremsstrahlung ($e^{-}(A,Z) \to e^{-}(A,Z) \nu \bar{\nu} $), neutron--nucleus bremsstrahlung ($n(A,Z) \to n(A,Z) \nu \bar{\nu} $), neutron--neutron bremsstrahlung ($n n \to n n \nu \bar{\nu} $), Cooper pairing of neutrons ($n n \to \nu \bar{\nu} $), among others. In the neutron star core, we quote just a few of all the possible reactions: baryon direct Urca ({\it e.g.} $p \, l \to n \nu_{e}$, $p \, l \to \Sigma \nu_{e}$), baryon modified Urca ({\it e.g.} $p B \, l \to n B \nu_{e}$, $p B \, l \to \Sigma B \nu_{e}$), baryon bremsstrahlung ({\it e.g.} $n n \to nn \nu \bar{\nu}$), lepton modified Urca ({\it e.g.} $e^{-} p \to \mu \, p \bar{\nu}_{\mu} \nu_{e}$) and Coulomb bremsstrahlung ({\it e.g.} $l \, p \to l \, \nu \bar{\nu}$).

Certainly, the emission of neutrinos is considered the main mechanism for the neutron star cooling~\cite{Sh96, Ya04}. In the analysis of this emission, the neutrino mean free path $\lambda$ is of central importance. Depending on the conditions of density and temperature, the neutrino mean free path ranges from small values compared with the neutron star radius, up to very large values. In the absence of magnetic field this has been extensively discussed in the literature (see for instance~\cite{tubbs,sawyer,iwamoto,backman,haensel87,Ho91,Re97,Re98,ReddyT,Re99,Bu98,Na99,Sh03,Ma03}). The neutrino mean free path tells us about the neutrino emissivity from the neutron star and therefore the degree of cooling of the compact object.

The addition of a strong magnetic field modifies these processes. Observational data on the magnetic field strength in the neutron star surface indicates that this magnitude varies within the range B$=10^{8}$--$10^{15}$G. A comprehensive and detailed review of the magnetic field in a neutron stars can be found in~\cite{Re03} and references therein. The magnetic field strength in the surface of a neutron star such as a young radiopulsar ($\tau \sim 10^{3}-10^{7}$yr) has values in the range B$=10^{11}$--$10^{13}$G. For an old radiopulsar ($\tau \sim 10^{8}-10^{10}$yr) this value decrease to B$=10^{8}$--$10^{9}$G, while also in the surface of a magnetar this value rise up to B$\sim 10^{15}$G and it can grow by several orders of magnitude in its dense interior~\cite{Du92}. The stability condition requiring that the total neutron star energy be negative leads to an upper bound on the magnetic field strength of B$\lesssim 10^{18}$G~\cite{Le77}.

The magnetic field establish a preferred axis for the neutron star, making the emission of neutrinos asymmetrical. This asymmetry has astrophysical implications and perhaps, the most important one is as a possible mechanism for the explanation of the ``pulsar kick problem": the observation that pulsars do not move with the velocity of its progenitor star, but rather with a substantially greater speed. Even thought this model has been objected as the only source to explain the problem pulsar kick (see for instance~\cite{Sa08}), an asymmetry of $\sim 1\%$ would be enough to understand this behavior~\cite{La98}. There are two main mechanisms responsible for this asymmetry. One is the effect of the magnetic field on the oscillation of the neutrinos~\cite{Ku96}. The second source of asymmetry are the parity violation reactions which take place inside the neutron star~\cite{La98,Bi93,Ch93,Bi97,Ho98,Ba99,Ch02,Sh05,Ka06,Ma11,Ma12,To19}. This last approach is the one that we adopt in this work.

In this work we analyze the asymmetry in the neutrino mean free path for the absorption reaction $\nu  + n \to e^{-}  + p$, in hot dense neutron matter. In a previous paper we have discussed the scattering process $\nu  + n \to \nu'  + n'$~\cite{To19}. By considering both reactions, the asymmetry in the neutrino emission can be originated from the differential cross section and from the neutrino mean free path (which is the inverse of the total cross section per unit volume). These two mechanism are independent and should be considered simultaneously to account for the actual asymmetric neutrino emission. While the first one is restricted to the scattering reaction and it gives us information on the way in which the weak interaction scatters the neutrinos, for the mean free path both reactions are present and it tells us about how often a neutrino interacts with a neutron. We consider that the mean free path is the relevant variable in this problem: if the mean free path is much larger than the size of the compact object itself, then the asymmetry in the differential cross section would not act, since it would be unlikely to have a collision.

As we have already mentioned, we consider the absorption reaction which takes place in hot dense neutron matter under a strong magnetic field. The magnetic field induces some degree of polarization of the system, which is partially responsible for the asymmetry in the mean free path. At this point, it is worth to mention that the total neutrino cross section shows a dependence on the angle of the incoming neutrino with respect to the magnetic field also in free space~\cite{Sh05}. Neutron matter is described using the Equation of State (EoS) developed in~\cite{Ag11,Ag14}. In this approach, we describe the nuclear interaction using the non--relativistic Skyrme potential model within a Hartree--Fock approximation. 

This work is organized as follows. In Section~\ref{nacs} we present the formalism for the neutrino mean free path. This is done in two sub--sections where we discuss the EoS in first place and then we give some details on the derivation of the cross section per unite volume. In the next step, we discuss our results in Section~\ref{Results}, where we also include the scattering mean free path previously evaluated. Finally, in Section~\ref{Summary} we give some conclusions.

%
\section{The Neutrino Absorption Cross Section}
\label{nacs}
In this section we present an expression for the neutrino absorption cross section in hot neutron matter under a strong constant magnetic field. Much of the information in this section have been already published in other works and we have done a summary of them for the convenience of the reader. But also because we add some specific information which should be given in the right context of our problem.

\begin{figure}[t]
\begin{center}
    \includegraphics[width = 0.45 \textwidth]{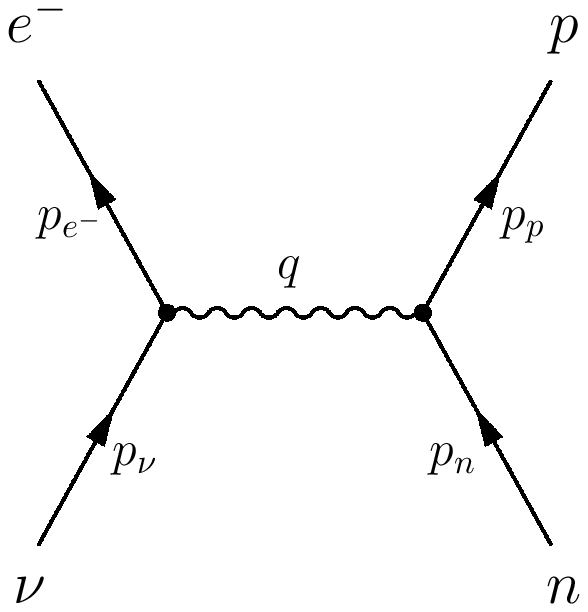}
\vskip 2mm
\caption{The lowest order Feynman diagram for the scattering reaction $\nu + n \to e^{-} + p$. The quantities $p_i$ and $q$ denote, respectively, the four--momentum of the involved particles and the
corresponding four--momentum transfer by the interaction.}
\label{figme1}
\end{center}
\end{figure}

The absorption reaction under consideration is the absorption of a neutrino by a neutron, having an electron and a proton as the final state,
\begin{equation}
\nu \, + \, n \, \rightarrow \, e^{-} \, + \, p,
\label{reacabs}
\end{equation}
where the Feynman diagram for this reaction is drawn in Fig.~\ref{figme1}. This reaction can take place either in free space or within a dense medium. We are considering pure hot non--relativistic neutron matter and to evaluate the cross section we need two basic elements: in first place, a model for the neutron matter. This means that we have to develop an Equation of State (EoS) for the dense medium under the influence of a strong magnetic field, from which we obtain the physical state of the system, characterized by the polarization, the single particle energies and the chemical potential for equilibrium. The second element is the evaluation of the diagram in Fig.~\ref{figme1} itself, using the standard rules for the evaluation of diagrams. In particular, we should employ a model for the weak--interaction which mediate this reaction. In two sub--section we address these points.

\subsection{The EoS model using a Skyrme interaction}
\label{eossi}
The EoS is evaluated using Hartree-Fock approximation with the Skyrme interaction~\cite{Ag11,Ag14}. We assume a system of neutrons within a strong magnetic field at finite temperature. The neutrons interact through the strong interaction among each other and with the external magnetic field. From the EoS, we obtain the degree of polarization of the system, the single particle energies of the neutrons and their chemical potential. This is done by giving the density of the system, its temperature and the intensity of the magnetic field, which we consider as a constant field in the $\hat{z}$--direction. This hypothesis on the magnetic field is not an important restriction as it should be employed locally (as well as the density and the temperature). For the whole neutron star one can implement a realistic model for the magnetic field. The curvature of such a field would allow us to consider it as locally uniform due to the scale of the neutrino--neutron absorption reaction.

Now we briefly describe how we obtain the different outcomes from the EoS. For a more detailed analysis we refer the reader to~\cite{Ag11,Ag14,To19}. The starting point is to define the adequate thermodynamical potential for our problem. For a system within a magnetic field ${\vec B}$, we employ,
\begin{equation}
{\cal U}={\cal F} - {\vec {\cal M}}\cdot {\vec B},
\label{tp}
\end{equation}
where ${\cal F}$ and ${\vec {\cal M}}$ are, respectively, the Helmhotz free energy density and the magnetization per unit volume of the system. The expression for the density of the system is given by,
\begin{equation}
\rho= \sum_{s_n=\pm 1} \frac{1}{(2 \pi)^{3}} \int d^{3} p_{n} f_{s_n}(E_{n},\mu_n, T).
\label{dent}
\end{equation}
Here $E_{n}$, $\mu_n$ and $T$ stands for the neutron single particle energy, its chemical potential and the temperature, respectively. The function $f_{s_n}(E_{n},\mu_n, T)$, in thermal equilibrium is given by the Fermi--Dirac particle distribution function,
\begin{equation}
f_{s_i}(E_{i},\mu_i, T)=\frac{1}{1+\exp[(E_{i}-\mu_i(T))/T]} \ .
\label{fdd0}
\end{equation}
It is straightforward to define the spin up and down partial densities as $\rho_{+}$ and $\rho_{-}$, respectively. We have $\rho=\rho_{+}+\rho_{-}$. The spin asymmetry is,
\begin{equation}
{\it A}= \frac{1}{\rho} \, \sum_{s_n=\pm 1} \frac{s_n}{(2 \pi)^{3}} \int d^{3} p_{n} f_{s_n}(E_{n},\mu_n, T),
\label{spinasym}
\end{equation}
or equivalently, ${\it A}=(\rho_{+}-\rho_{+})/(\rho_{+}+\rho_{+})$. At this point it is convenient to give the expression for the neutron single particle energy, $E_n$. Using the Hartree--Fock model with the Skyrme interaction, we have~\cite{Ag11,Ag14},
\begin{equation}
E_n = m_n \, + \, \frac{p_n^2}{2m^{*}_{s_n}} - s_n \mu_{Bn} B\, + \, \frac{v_{s_n}}{8},
\label{speexp}
\end{equation}
where $\mu_{Bn}=-1.913 \mu_N$ is the
anomalous magnetic moment of the neutron in units of the nuclear magneton $\mu_N$. The potential term, $v_{s_n}$ depends on the density, the temperature and the magnetic field, but not on the momentum and it is given by,
\begin{equation}
v_{s_n}=a_0 (1- s_n {\it A}) \rho+2 (b_0+s_n b_1) \, {\cal K}_{s_n=1},
\label{SkmPot}
\end{equation}
where,
\begin{equation}
{\cal K}_{s_n} = \frac{1}{(2 \pi)^{3}} \, \int \, d^{3} p \; p^{2} f_{s_n}(E_{n},\mu_n, T) \, .
\label{SkmKIn}
\end{equation}
The constants $a_0=4 t_0 (1-x_0) +2 t_3 \rho^{\sigma} (1-x_3)/3$, $b_0=t_1(1 - x_1)+ 3 t_2 (1 + x_2)$ and $b_1=- t_1(1 - x_1)+t_2 (1 + x_2/2) $, are written in terms of the standard parameters of the Skyrme model, $t_0$, $t_1$, $t_2$, $x_0$, $x_1$, $x_2$ and $\sigma$. In Eq.~(\ref{speexp}), for the effective mass we have,
\begin{equation}
\frac{1}{m^*_{s}}=\frac{1}{m_n}+\frac{1}{4}\, \rho \, ( b_0 + s \, b_1 \, {\it A} ) .
\label{SkmEffMass}
\end{equation}

The chemical potential corresponding to the physical state, does not depend on the spin projection of the neutron due to the minimization process. To see this point we write,
\begin{equation}
\mu_{s_n}=\frac{\partial {\cal U}}{\partial \rho_{s_n}}.
\label{e:cp}
\end{equation}
This expression can be rewritten in terms of the spin asymmetry ${\it A}$, as,
\begin{equation}
\mu_{s_n} = \frac{\partial {\cal U}}{\partial \rho }+s_n \left(\frac{1-s_n {\it A}}{\rho}\right) \frac{\partial {\cal U}}{\partial {\it A}}.
 \label{cp2}
\end{equation}
The difference between the two chemical potentials is then,
\begin{equation}
\mu_+ - \mu_- =\frac{2}{\rho}\frac{\partial {\cal U}}{\partial {\it A}} \,
\label{cp3}
\end{equation}
which shows, that the minimization of ${\cal U}$ with respect to {\it A} implies the existence of a unique chemical potential in the physical state. We should emphasized that this minimization is performed with the constrain of a fixed density. This is a self--consistent process: we need $\mu_{n}$ to evaluate $\rho_{+}$ and $\rho_{-}$, which defines the spin asymmetry ${\it A}$, needed in the single particle energy, etc. Summarizing, given the density, temperature and the magnetic field of the system, from the EoS we obtain the actual physical state, characterized by the chemical potential, the single particle energies of the neutrons and the spin asymmetry which is a global property of the system.

\subsection{The Absorption Neutrino Cross Section for a polarized system}
\label{ncspol}
In this sub--section we show an expression for the absorption neutrino cross section per unite volume for a polarized system. The formalism of this sub--section is taken from the work of Arras and Lai~\cite{Ar99}, where the reader can find a complete derivation. We briefly summarized some elements for convenience and we also add some particular expressions not given in~\cite{Ar99}, that we need in our work.

The aim of this sub--section is to write an analytical expression for the mean free path for the absorption reaction $\nu  +  n  \to  e^{-} + p$, depicted in~Fig.~\ref{figme1}. From this diagram, the weak interaction is given by the effective Hamiltonian,
\begin{equation}
{\cal H}_{int} = \frac{G_F}{\sqrt{2}} \, \bar{\Psi}_{p} \gamma_{\mu} (g_V-g_A \gamma_5) \Psi_{n} \, \bar{\Psi}_{e} \gamma^{\mu} (1- \gamma_5) \Psi_{\nu} \, + \, H.c.
\label{effH}
\end{equation}
Here $G_F$ is the Fermi weak coupling constant ($G_F/(\hbar c)^{3}=1.16637(1) \times 10^{-5}$~GeV$^{-2}$). For the vector and axial--vector couplings we have $g_V=0.973$ and $g_A=1.197$, respectively. The total absorption cross section per unit volume can be written as,
\begin{equation}
\frac{\sigma_{abs}}{V} = \int \, d \Pi_{p} \, d \Pi_{e} \, d \Pi_{n} \, {\cal W}^{abs}_{fi} \,  (1-f_{s_p}(E_{p},\mu_p, T)) (1-f_{s_e}(E_{e},\mu_e, T)) f_{s_n}(E_{n},\mu_n, T).
 \label{cs1}
\end{equation}
In this expression ${\cal W}^{abs}_{fi}$ is the transition rate, which is linked to the Hamiltonian through the $S$--matrix. The $S$--matrix is defined as,
\begin{equation}
S_{fi}= \imath \,\int \, d^{4}x \,  {\cal H}_{int}.
\label{smatrix}
\end{equation}
The square of $S_{fi}$, divided by time is the transition rate:
\begin{equation}
{\cal W}^{abs}_{fi} = \frac{\mid S_{fi} \mid^{\, 2}}{t}.
\label{tranW}
\end{equation}
The Eq.~(\ref{cs1}), is in fact the Fermi Golden Rule, where we sum over final states and average over the initial ones, if we do not know the initial state. We are considering massless neutrinos which are left-handed (or polarized), we assign also a value for its momentum and direction. That is, we know the initial state for the neutrino. In the same equation, the $\int \, d \Pi_{N}$ represents the state summation for the particle $N$. Is is convenient to show the explicit expression for each particle, together the corresponding single particle energy.

Protons and electrons are charged particles and therefore, their energy levels are partially quantized according to the Landau levels. In particular, the single particle energy for a proton which interacts only with a constant magnetic field (which we take as the $\hat{z}$--direction) is,
\begin{equation}
E_{p}=m_p+\frac{p^{2}_{\, p,z}}{2 m_{p}}+\frac{e B}{m_p} \, (N_p+\frac{1}{2}) \, - \,s_p \mu_{Bp} B ,
\label{spprot0}
\end{equation}
where $\mu_{Bp}=2.793 \mu_N$ and $N_p=0$, 1, 2, ... is the energy level quantum number for the proton Landau state. The quantization axis for a charge particle is perpendicular to the magnetic field--direction. For the proton state summation we have,
\begin{equation}
\int  \, d \Pi_{p} \, = \sum^{N_{p, \, max}}_{N_p=0} \,\sum^{R_{p, \, max}}_{R_p=0} \, \sum_{s_p=\pm 1}  \int^{\infty}_{-\infty} d p_{p, \, z} \; \frac{L \, p_{p, \, z}}{2 \pi},
\label{sumsn}
\end{equation}
where $N_{p, \, max}$ is determined by the conservation of energy and $R_{p}$ is the quantum number for the proton guiding center, where the cutoff $R_{p, \, max} \simeq e B {\cal A}/2 \pi$ (the degeneracy of the Landau level) limits the guiding center to lie within the normalization volume $V= {\cal A} L$, where $L$ is the length along the $\hat{z}$--axis and ${\cal A}$ is the area.

Due to the small mass of the electron, we have employed the relativistic expression for the energy,
\begin{equation}
E_{e}=(m^{2}_{e} + 2 e B N_e + p^{2}_{e, \, z})^{1/2},
\label{spee0}
\end{equation}
where $N_e=0$, 1, 2, ... is the energy level quantum number for the electron Landau state. The particular value for the magnetic moment for the electron allows us to employ a single index ($N_e$) in its energy. To specify the quantum state of the electron we need also $\sigma_{e}=\pm 1$, the spin projection along ${\bf \Pi=}{\bf p}_e+ e {\bf A}$ and $R_e=$0, 1, 2, ... which plays the same role as for the proton. For a detail discussion on the solution of the Dirac equation for the electron we refer the reader to~\cite{So68}. The summation for the electron is then,
\begin{equation}
\int  \, d \Pi_{e} \, = \sum^{N_{e, \, max}}_{N_e=0} \, \sum_{\sigma_e=\pm 1} \, c(N_e,\sigma_e) \,\sum^{R_{e, \, max}}_{R_e=0}   \int^{\infty}_{-\infty} d p_{e, \, z} \; \frac{L \, p_{e, \, z}}{2 \pi},
\label{sumsn2}
\end{equation}
where the function $c(N_e,\sigma_e)=1-\delta_{N_e, \, 0} \delta_{\sigma_{e},-\sigma_{e 0}}$, with $\sigma_{e 0}=-$sgn$(p_{e, \, z})$. This function is equal to one, except for its null value when $N_e=0$ and $\sigma_{e}=-\sigma_{e 0}$. This is needed because for the ground Landau level, the electron spin is opposite to the magnetic field. This means that we can only have the spin projection $\sigma_{e 0}$~\footnote{Just for clarity, we simplify this point by given the non--relativistic limit. In this case, we have $N_e=n+1/2+\tilde{s_e}$, with $n=0$, 1, 2, ... and $\tilde{s_e}=\pm1/2$. First we consider the ground state $N_e=0$, therefore $n=0$ and $\tilde{s_e}=-1/2$, having only one possible spin projection. If we take a fixed value for $N_e$, with the condition that it is not equal to zero, then we can have the two spin projections: $N_e=n+1/2+\mid\tilde{s_e}\mid$ and $N_e=n'+1/2-\mid\tilde{s_e}\mid$, with $n'=n+1$.}. The cutoff $R_{e, \, max}$ takes the same value as for the proton.

The single particle energies in Eqs.~(\ref{spprot0}) and (\ref{spee0}), are the ones employ in this work, as we are considering pure neutron matter. Once the neutrino is absorbed by the neutron, the final proton and electron do not find others fermions of the same kind. In this sense, in Eq.~(\ref{cs1}), we should make the replacement $(1-f_{s_p}(E_{p},\mu_p, T)) (1-f_{s_e}(E_{e},\mu_e, T)) \rightarrow 1$. However, we will retain these functions to preserve a more general expression.

Finally, for the neutron, we have,
\begin{equation}
\int  \, d \Pi_{n} \, = \sum_{s_n=\pm 1} \frac{1}{(2 \pi)^{3}} \int d^{3} p_{n}.
\label{sumsn3}
\end{equation}
The single particle energy for the neutron has been already given in Eq.~(\ref{speexp}). As we have mentioned, in Eq.~(\ref{cs1}), we sum over all possible final states and we average over the initial ones. The next step is to insert all wave functions into this equation and obtain the final expression for the neutrino cross section. This procedure is developed in full detail in~\cite{Ar99}, we will not repeat it here. We employ a non--relativistic wave function for neutron. In this point we are interested in the spin term of this wave function. In unpolarized matter, one makes an average over the spin up and down contributions, $\ket{u}$ and $\ket{d}$, respectively. For polarized matter, we employ a single mixed spin wave function $\ket{\chi_{n}}$ (for details see Appendix~B in~\cite{To19}),
\begin{equation}
\ket{\chi_{n}}  =   \sqrt{\frac{1+{\it A}}{2}} \, \ket{u} + \sqrt{\frac{1-{\it A}}{2}} \, \ket{d},
\label{spinwf}
\end{equation}
where ${\it A}$ is the spin asymmetry as defined in Eq.~(\ref{spinasym}). The mean value of the spin projection operator $\hat{S}_{z}$, using this wave function is,
\begin{equation}
\bra{\chi_{n}} \hat{S}_{z} \ket{\chi_{n}}={\it A} \, \frac{\hbar}{2} \ ,
\label{spinwfmv}
\end{equation}
which is the same as the mean value of the spin projection operator for the whole system, as required by the mean value for a mixed wave function~\cite{Cohen}. In what follows, we employ the neutron spin wave function in Eq.~(\ref{spinwf}), for the evaluation of the cross section.

We give now the expression for the cross section. To do so, one has to replace each particle wave function in Eq.~(\ref{cs1}). As mentioned, the procedure is developed in detail in the~\cite{Ar99}. With the addition of the neutron spin wave function it is obtained, 
\begin{eqnarray}
\frac{\sigma_{abs}}{V} & = & \frac{G^{2}_F}{2} \, \frac{e B}{2 \pi} \, \sum^{N^{max}_{e}}_{N_e=0} \, \int^{\infty}_{-\infty} \frac{d \, p_{e, \, z}}{2 \pi} (1-f_{N_e}(E_{e},\mu_e, T))
\, \sum^{N^{max}_{p}}_{N_p=0} \, \int^{\infty}_{-\infty} \frac{d^{\, 2} p_{n, \, \perp}}{(2 \pi)^{2}} \, \sum_{s_p=\pm 1} \nonumber \\
& \times &  \Biggl( \biggl(\frac{1+{\it A}}{2} \biggr)  S_{s_p,1,N_p,N_e} L_{\mu\nu} N^{\mu\nu} \mid_{s_p,1}
+  \biggl( \frac{1-{\it A}}{2}\biggr)  S_{s_p,-1,N_p,N_e} L_{\mu\nu} N^{\mu\nu} \mid_{s_p,-1} \Biggr),
\label{cs2}
\end{eqnarray}
where $p_{\, n,\perp} = \sqrt{p^{2}_{\, n,x}+p^{2}_{\, n,y}}$. In this expression $L_{\mu\nu}$ and $ N^{\mu\nu}$ are the leptonic and hadronic tensors, respectively, as defined in Eqs.~(D12) and (D13) in~\cite{Ar99}.  We have introduced the structure function for the absorption process as,
\begin{eqnarray}
S_{s_p,s_n,N_p,N_e} & = & \int^{\infty}_{-\infty} \, \frac{d p_{\, n,z}}{2 \pi} \, \int^{\infty}_{-\infty} \, \frac{d p_{\, p,z}}{2 \pi} \,
(2 \pi)^{2} \, \delta(E_e+E_p-|p_{\nu}|-E_n) \nonumber \\
&\times& \delta(p_{\, e,z}+p_{\, p,z}-p_{\, \nu,z}-p_{\, n,z}) \; f_{s_n}(E_n,\mu_n,T) \, (1-f_{s_p}(E_p,\mu_p,T)).
\label{stfun0}
\end{eqnarray}
An analytical expression for this function is given in the Appendix~\ref{strucfunc}, where at variance with~\cite{Ar99}, this function is evaluated in the case where $m_p \neq m_n$. Finally, the contraction of the leptonic and hadronic currents are given by,
\begin{eqnarray}
L_{\mu\nu} N^{\mu\nu}\mid_{s_p,s_n}(N_e=0)&=& \theta(p_{e, \,z}) \,
I^2_{0,N_p }(t) \,
\biggl( g_V^2 +3 g_A^2
+\left(g_V^2 - g_A^2\right) \cos(\theta_{\nu}) \nonumber\\
&&+2 g_A \left(g_A + g_V\right)\left( s_p+s_n  \cos(\theta_{\nu}) \right)
-2g_A\left(g_A-g_V \right)\left( s_n+s_p \cos(\theta_{\nu})\right)\nonumber\\
&& + \left(
g_V^2-g_A^2
+\left(g_V^2+3g_A^2\right)\cos(\theta_{\nu})
\right) \, s_n s_p
\biggr),
\label{trazaNe0}
\end{eqnarray}
where $\theta_{\nu}$ is the angle among the neutrino and the magnetic field and the function $I_{N_{e},N_p }(t)$ is given by,
\begin{equation}
I_{N_{e},N_p }(t) = \biggl( \frac{N_p!}{N_e!} \biggr)^{1/2} \exp^{-t/2} t^{(N_e-N_p)/2} {\cal L}^{N_e-N_p}_{N_p}(t),
\label{funcI}
\end{equation}
where $t=\omega^{2}_{\perp}/2 e B$ and for the definition of the Laguerre polynomials ${\cal L}^{i}_{j}$, we have adopted the one from~\cite{Ab72}.
When $N_e \geq 1$, we have,
\begin{eqnarray}
L_{\mu\nu} N^{\mu\nu}\mid_{s_p,s_n}(N_e \geq 1)&=&
g_V^2 \,
  \biggl( I^2_{N_{e}-1,N_p }(t) \Sigma^{-}_{N_e}(p_{e, \,z}) + I^2_{N_{e},N_p }(t) \Sigma^{+}_{N_e}(p_{e, \,z}) \biggr) (1+s_n s_p)\nonumber\\
 &+&g_A^2 \,
  \biggl( I^2_{N_{e}-1,N_p }(t) \Sigma^{-}_{N_e}(p_{e, \,z}) \, (3+\cos(\theta_{\nu})+2 (s_n-s_p) (1+\cos(\theta_{\nu}))\nonumber\\
&&-s_n s_p (1+ 3 \cos(\theta_{\nu})))+I^2_{N_{e},N_p }(t) \Sigma^{+}_{N_e}(p_{e, \,z}) \, (3-\cos(\theta_{\nu})\nonumber\\
&&-2 (s_n-s_p) (1-\cos(\theta_{\nu}))-s_n s_p (1+ 3 \cos(\theta_{\nu})))\biggr)\nonumber\\
&+& 2 g_V g_A \, \biggl(I^2_{N_{e}-1,N_p }(t) \Sigma^{-}_{N_e}(p_{e, \,z}) (-1+\cos(\theta_{\nu}))\nonumber\\
&&+ I^2_{N_{e},N_p }(t) \Sigma^{+}_{N_e}(p_{e, \,z}) (1+\cos(\theta_{\nu}))\biggr) (s_n + s_p),
\label{trazaNeNp}
\end{eqnarray}
where,
\begin{equation}
\Sigma^{\pm}_{N_e}(p_{e, \,z}) \equiv \frac{1}{2} \, \biggl( 1 \pm \frac{p_{e, \,z}}{\mid (p^{2}_{e, \,z}+2 e B N_e)^{1/2} \mid} \biggr).
\label{sigmael}
\end{equation}
The expression in Eq.~(\ref{trazaNe0}), is the same as the one in~\cite{Ar99}. But for the one in Eq.~(\ref{trazaNeNp}), we have considered all spin terms.

Note that the neutrino mean free path is obtained from the cross section as $\lambda_{abs}=(\sigma_{abs}/V)^{-1}$. In the next section we discuss our results.

\section{Results and discussion}
\label{Results}
We present now our results for the neutrino mean free path in homogeneous hot neutron matter under the presence of a strong magnetic field. We consider a range of densities of $0.04 \leq\rho \leq 0.4$ fm$^{-3}$, corresponding approximately to the outer core region a neutron star, temperatures up to T=30 MeV and different values of the magnetic field intensity ranging from B=$0$ up to B=$10^{18}$ G. The EoS is evaluated within the Hartree--Fock model, using LNS Skyrme interaction developed by Cao {\it et al.}~\cite{Ca06}. We have developed our formalism assuming a particular form for the single--particle energy for the neutron, which is the one from the Skyrme model. This expression is shown in Eq.~(\ref{speexp}). In~\cite{To19} we have employed the same model together with the Brueckner--Hartree--Fock (BHF) approach using the Argonne V18 \cite{argonne} nucleon-nucleon potential supplemented with the Urbana IX \cite{urbana} three-nucleon force. In that work we have obtained a good agreement between both models for the inelastic dispersion of neutrinos by neutrons. Note that the LNS Skyrme interaction is specially suitable for a comparison with the BHF--model, since its parameters were determined by fitting the nuclear matter EoS calculated in the BHF framework.

Before the discussion of our results, it is convenient to make a summary of the spin asymmetry of the system, which have been already analyzed in~\cite{To19} for the same interaction (see in particular the Fig.~3). The spin asymmetry ${\it A}$ characterizes the degree of polarization of the system. That is, we consider a system of neutrons interacting with each other through the strong interaction and with an external strong magnetic field. The strong interaction favors an equal number of neutrons with spin up and down ({\it i.e. A=}0), while the magnetic field tries to align all the neutron spins antiparallel to it ({\it i.e. A=}-1). The actual value for ${\it A}$ is then obtained through an energy minimization calculation from the EoS, as discussed in sub-Section~\ref{eossi}. As expected, the magnitude of {\it A} increases for decreasing densities and also for growing values of the magnetic field. In fact, within the range of B from $10^{14}$G up to $2.5 \times 10^{18}$G, we have $\log_{10}(\mid A \mid) \cong a \log_{10}(B)+b$, where $a \cong 1$ and $b$ is approximately constant for a fixed value of the density (this behavior is depicted in panels $b)$ and $d)$ in Fig.~3, in~\cite{To19}). Our concern is the neutrino mean free path and the corresponding cross section has different values according to the state of polarization of the neutron matter. This is developed in the following lines.

We turn now to the analysis of the absorption structure function as defined in Eq.~(\ref{stfun0}). An analytical expression for this structure function is given in the Appendix~\ref{strucfunc}. At variance with the well studied structure function for the dispersion mean free path (see Eq.~(23) in~\cite{To19}), this structure function has some particular features which deserves to be discussed. One should keep in mind that our structure function represents only a fraction of the proton--neutron phase space and due to this, it depends on many variables. Beyond its rather simple expression, it is the great number of independent variables which makes it difficult to analyze. Following the same pattern as for the dispersion structure function, we plot the absorption structure function as a function of $q_0$ (the energy transfer by the weak interaction). But instead of using a fix value for $q_z$, we employ $q_z \cong q_0 - |\vec p_{\nu}| (1-\cos(\theta_{\nu}))$. This expression is obtained by solving the first two expressions in Eqs.~(\ref{enemc}), for $q_z>0$ and $E_e \cong p_{\, e,z}$.

\begin{figure}[t]
\begin{center}
\vskip -4.5cm
    \includegraphics[scale=0.47]{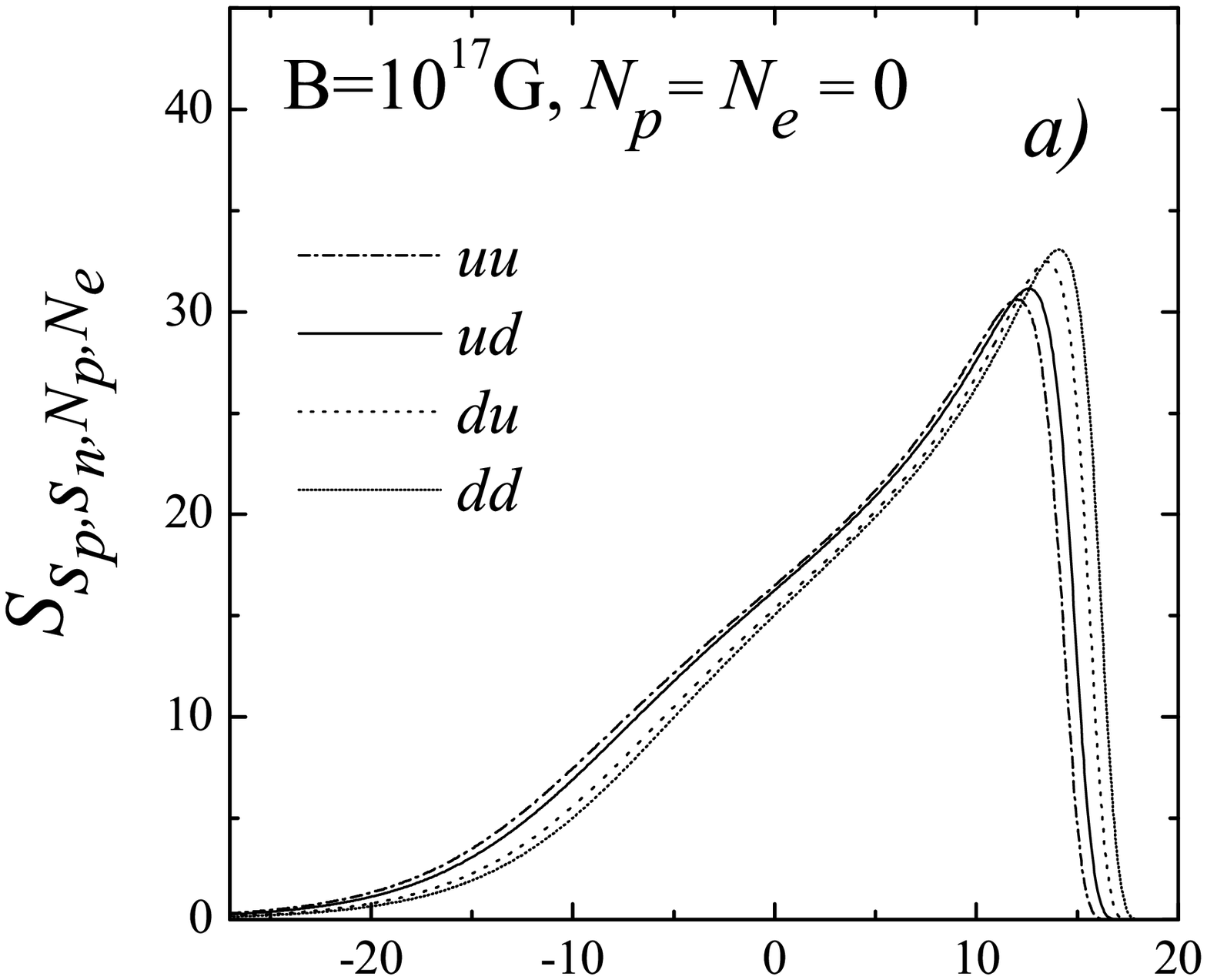}
\vskip -6cm
    \includegraphics[scale=0.47]{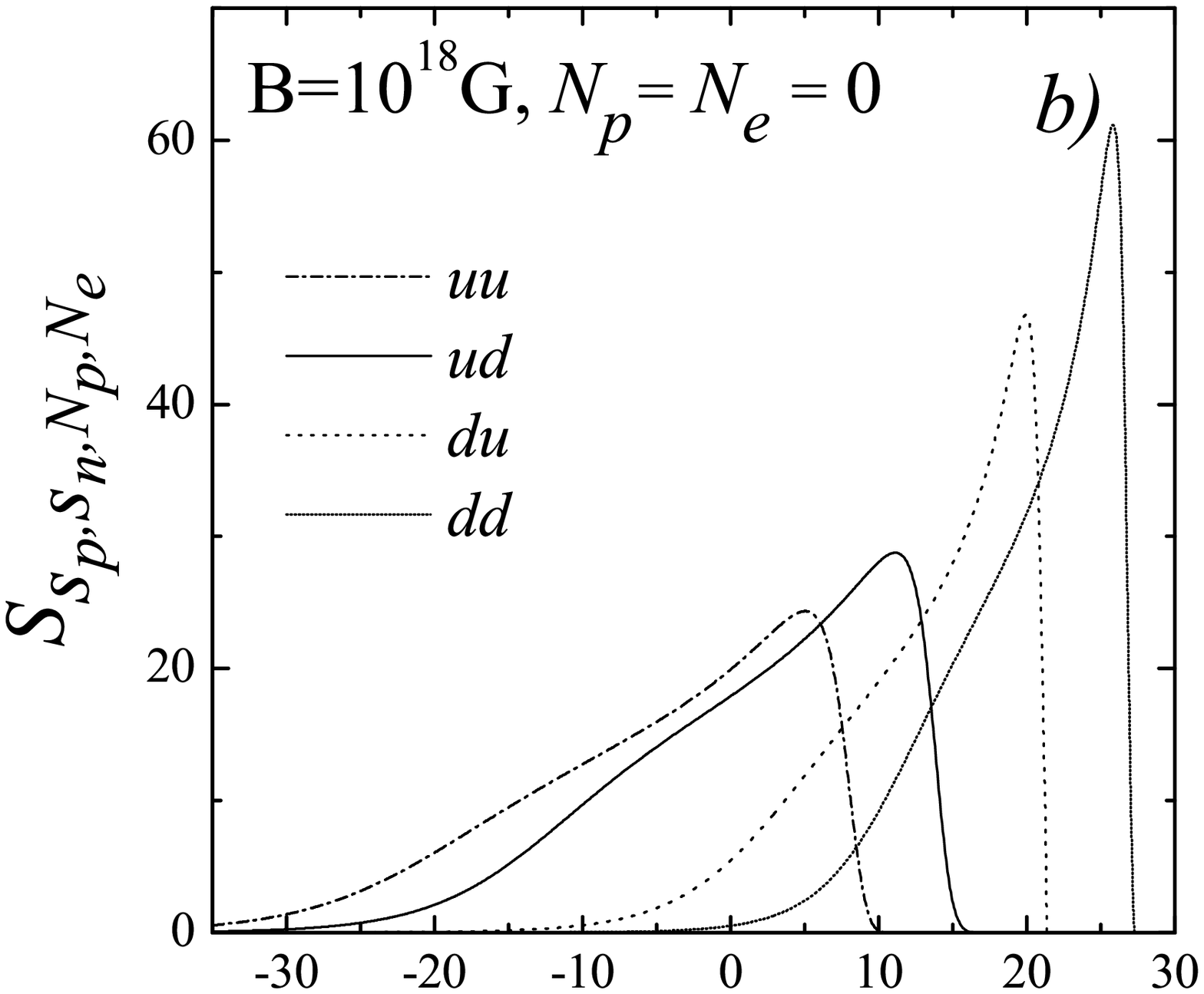}
\caption{Energy dependence of the structure function ${\cal S}_{s_p,s_n,N_p,N_e}$ for $\rho=0.16$ fm$^{-3}$.  In all panels we consider $N_p=N_e=0$, we take $q_z>0$ and $q_z \cong q_0 - |\vec p_{\nu}| (1-\cos(\theta_{\nu}))$ and we employ as a representative value for the square of the transverse momentum transfer by the neutron, $p_{n, \, \perp}=170$MeV. Also we use $|\vec p_{\nu}| = 3 T$, with $T=15$MeV. The values for $s_p,s_n$ are $uu$, $ud$, $du$ and $dd$. In panels $a)$ and $b)$ we show results for two values of the the magnetic field intensity, where we have used $\theta_{\nu}=0$.}
\label{figme2}
\end{center}
\end{figure}

In Fig.~\ref{figme2} we plot the structure function at a fixed density for different proton--neutron spin projections, denoted as $s_p,s_n=\, uu, \, ud, \,du$ and $dd$. In this figure we study the effect of the magnetic field over the structure function. The first obvious result is that the split among the different spin components is more relevant for $B=10^{18}$G. This split is due to two main elements. In first place, to the coupling of the magnetic field with the magnetic moment of protons and neutrons: from Eqs.~(\ref{spepn}) we notice that there is an energy shift of $\Delta E=(s_p \mu_{Bp}-s_n \mu_{Bn}) B$. Keeping in mind that $\mu_{Bp}>0$ and $\mu_{Bn}<0$, the main source of the split is understood. Secondly, the neutron effective mass depends on its spin projection, which represents the second contribution to the split. However, due to the particular Skyrme model that we have employed, this effect is small.

The shape of the different structure functions is linked to the single--particles energies and to the chemical potential derived from the EoS. But it is the area under the different functions which really matters: comparing the different areas, the bigger ones leads to bigger cross sections and smaller mean free paths. Let us call the different areas under each structure functions as $\int S_{s_p,s_n,N_p,N_e}$. From Fig.~\ref{figme2} and assuming that the whole area contributes to the cross section, we notice that $\int S_{uu,0,0} < \int S_{ud,0,0}$ and $\int S_{du,0,0} < \int S_{dd,0,0}$: for a fix proton spin projection, the contribution for neutrons with spin up is smaller than the one with spin down. The same behavior takes place for the dispersion structure function, having the same origin, which is the character of the phase space for polarization matter: the phase space for neutrons with spin up is smaller than that of neutrons with spin down. A complete discussion on this point is given in~\cite{To19}. A corollary of this discussion is that the structure function has a clear spin dependence and in the spin summation in Eq.~(\ref{cs2}), it can not be taken as a common factor.

\begin{figure}[t]
\begin{center}
    \includegraphics[scale=0.53]{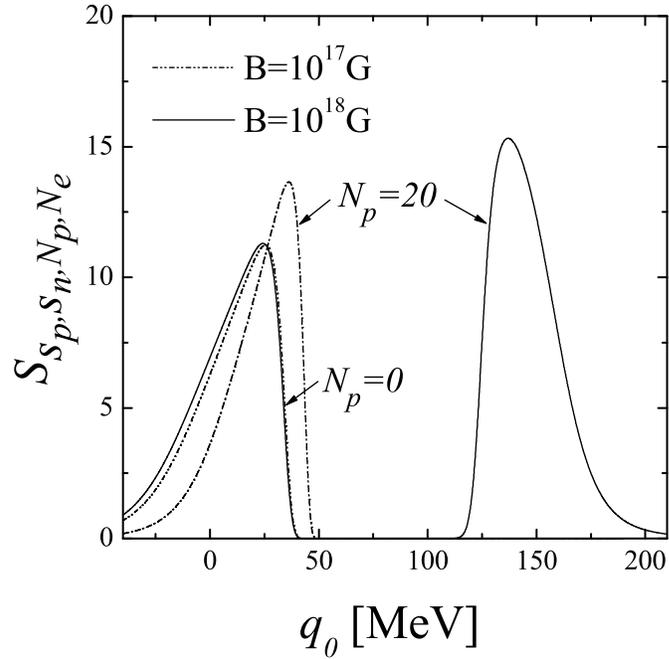}
\vskip 2mm
\caption{Dependence of the structure function ${\cal S}_{s_p,s_n,N_p,N_e}$ with $N_p$ for two values of the magnetic field intensity and $N_e=0$. We have considered $\theta_{\nu}=\pi/2$, $p_{n, \, \perp}=120$MeV, $|\vec p_{\nu}| = 3 T$, with $T=30$MeV and $s_p,s_n=ud$, while the others conditions are the same as in Fig.~\ref{figme2}.}
\label{figme3}
\end{center}
\end{figure}

In the following two figures, we limit ourselves to one spin configuration for simplicity, as the other contributions have the same behavior. In Fig.~\ref{figme3}, we show the $N_p$--dependence of the structure function for two values of the magnetic field. By drawing the structure function for $N_p=0$ and for $N_p=20$, we observe an energy shift stemming from the $(N_p+1/2) \, e B/m_p$--term in the proton single particle energy. Being this term proportional to $B$, it is straightforward to understand that this shift is one order of magnitude bigger for $B=10^{18}$G than for $B=10^{17}$G. The problem here is up to which $N_p$--value should we sum up. Or equivalently, which is the biggest value for $q_0$. The value for $q_0$ is limited by the conservation of energy: $|p_{\nu}|+E_n=E_e+E_p$. The initial energy of the system depends on the particular values of the momentum carried by each particle, its potential energy and the value for the magnetic field. Note that $q_0=|p_{\nu}| - E_{e}$: $|p_{\nu}|$ has a fixed value and $E_e < |p_{\nu}|+E_n$. This gives a maximum value for $q_0$, keeping in mind that $E_p \cong 0$ is not a realistic situation. This is the first constrain to the value for $N_p$, but as we discuss soon $N_p$ has also restrictions by the accessible phase space. Before ending this paragraph, it is worth to mention that $N_e$ is indirectly present in $q_0$. This quantum number is part of the electron single particle energy. In this figure, we have employed the approximate equality, $q_z \cong q_0 - |\vec p_{\nu}| (1-\cos(\theta_{\nu}))$, which is valid only if $N_e=0$. A similar figure can be done for $N_e \neq 0$, but leading to the same conclusions.

Coming back the the energy conservation $E_e+E_p=|p_{\nu}|+E_n$, it is convenient to make some comment on the relative values for $N_p$ and $N_e$. We show a simple model to compare the $N_p$--contribution to the proton energy term $\Delta E_{Np} \equiv N_p e B/m_p$ with the corresponding term for the electron, $\Delta E_{Ne} \equiv (m^{2}_{e} + 2 e B N_e)^{1/2}-m_{e}$. By defining $\Delta E_{tot} \equiv \Delta E_{Np}+\Delta E_{Ne}$ and just to give an example, we set the maximum possible value for $\Delta E_{tot}$, at $\Delta E^{max}_{tot}=64$MeV. We consider two cases: $i)$ B$=10^{18}$G, we have $\Delta E_{Np=10}=63$MeV, but $\Delta E_{Ne=1}=108$MeV, which means that no electron Landau level contributes to the cross section and we have to sum $N_p$ from zero up to ten. $ii)$ B$=10^{17}$G, $\Delta E_{Np=100}=63$MeV and in this case, $\Delta E_{Ne=3}=59$MeV. Then, we have combinations among the proton and the electron Landau levels: $N_p=100$ and $N_e=0$, $N_p=0$ and $N_e=3$, $N_p=1$ and $N_e=1$, etc. Due to the small electron mass, the energy gap is always bigger for the electron.

\begin{figure}[t]
\begin{center}
    \includegraphics[scale=0.53]{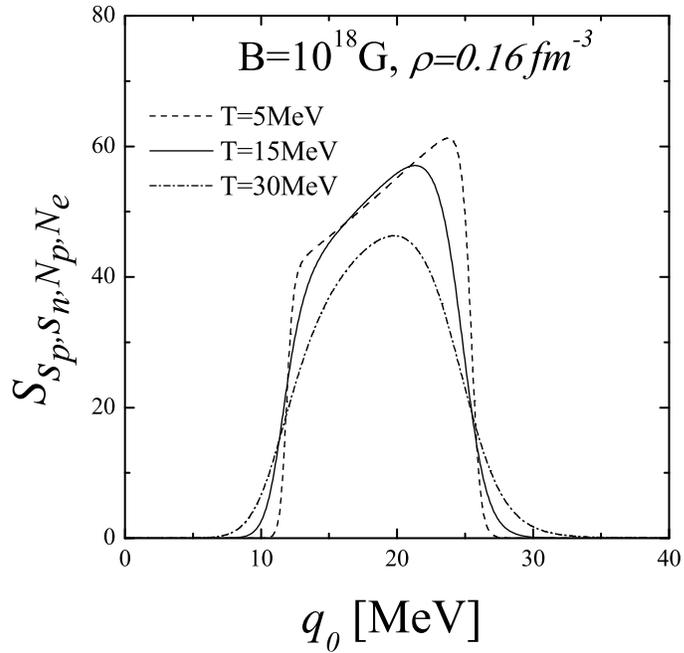}
\vskip 2mm
\caption{The structure function ${\cal S}_{s_p,s_n,N_p,N_e}$ for different temperatures. We have considered $\theta_{\nu}=0$, $p_{n, \, \perp}=70$MeV, $|\vec p_{\nu}| = 3 T$, with $T=30$MeV  and $s_p,s_n=ud$, while the others conditions are the same as in Fig.~\ref{figme2}.}
\label{figme4}
\end{center}
\end{figure}

In Fig.~\ref{figme4}, which is the last one for the structure function, we consider the temperature dependence of this function for $B=10^{18}$G and for three values of the temperature T$=5$, $15$ and $30$MeV. By comparing this results with the ones from the dispersion structure function (see Fig.~4a in~\cite{To19}), we notice that the behavior of the absorption structure function with temperature is quite different from the one in the dispersion process. For the dispersion process, the area  under the structure function strongly grows with temperature. At variance, for the absorption one, the areas are similar,  but with a clear decrease as one  increases the temperature. The absorption structure function represents only a part of the available phase space and as so, gives a different result. A complete analysis of the temperature dependence requires the full phase space of the problem. This is done soon, when we discuss the temperature dependence of the neutrino mean free path.

\begin{figure}[t]
\begin{center}
\vskip -30mm
    \includegraphics[scale=0.47]{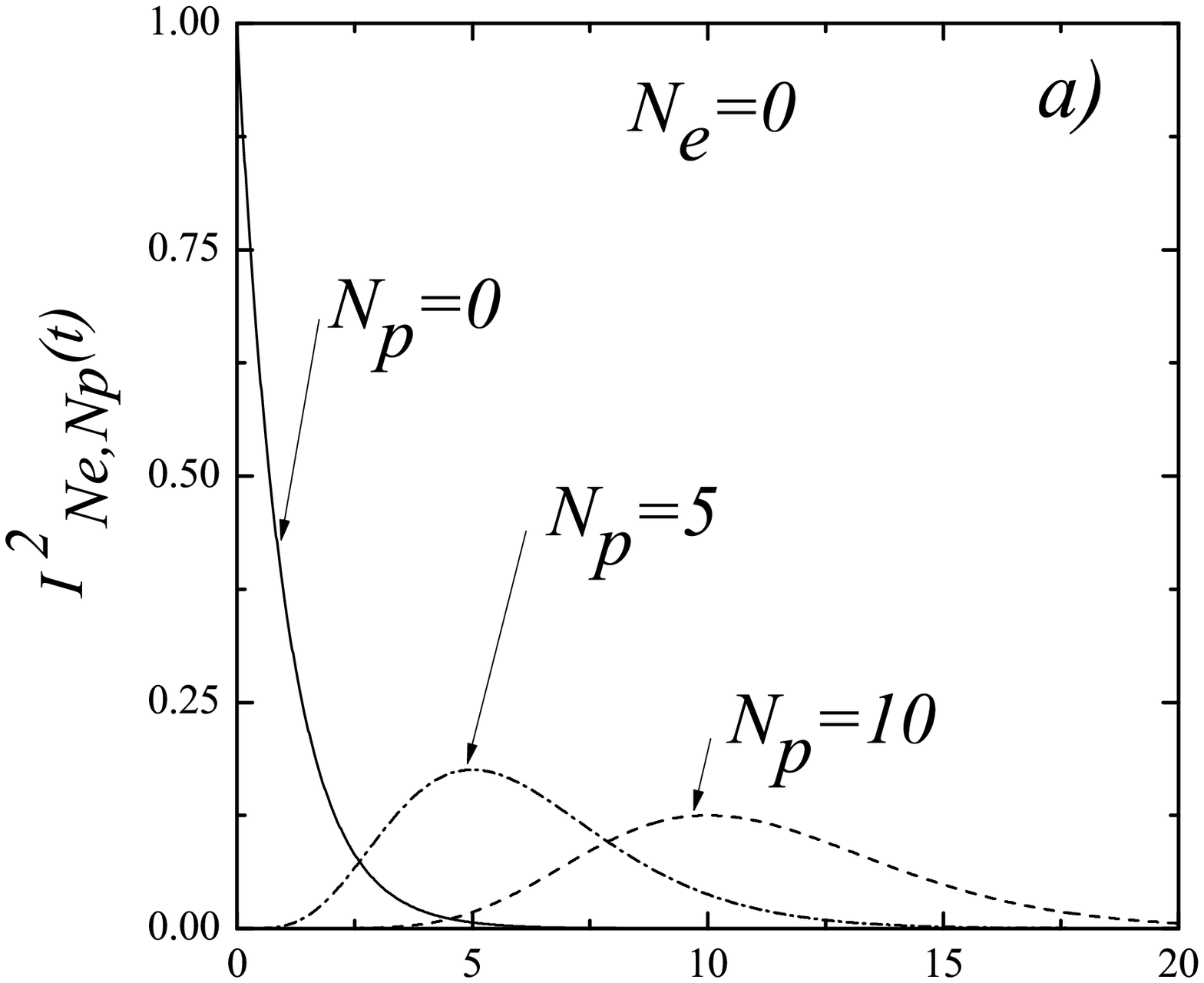}
\vskip -6cm
    \includegraphics[scale=0.47]{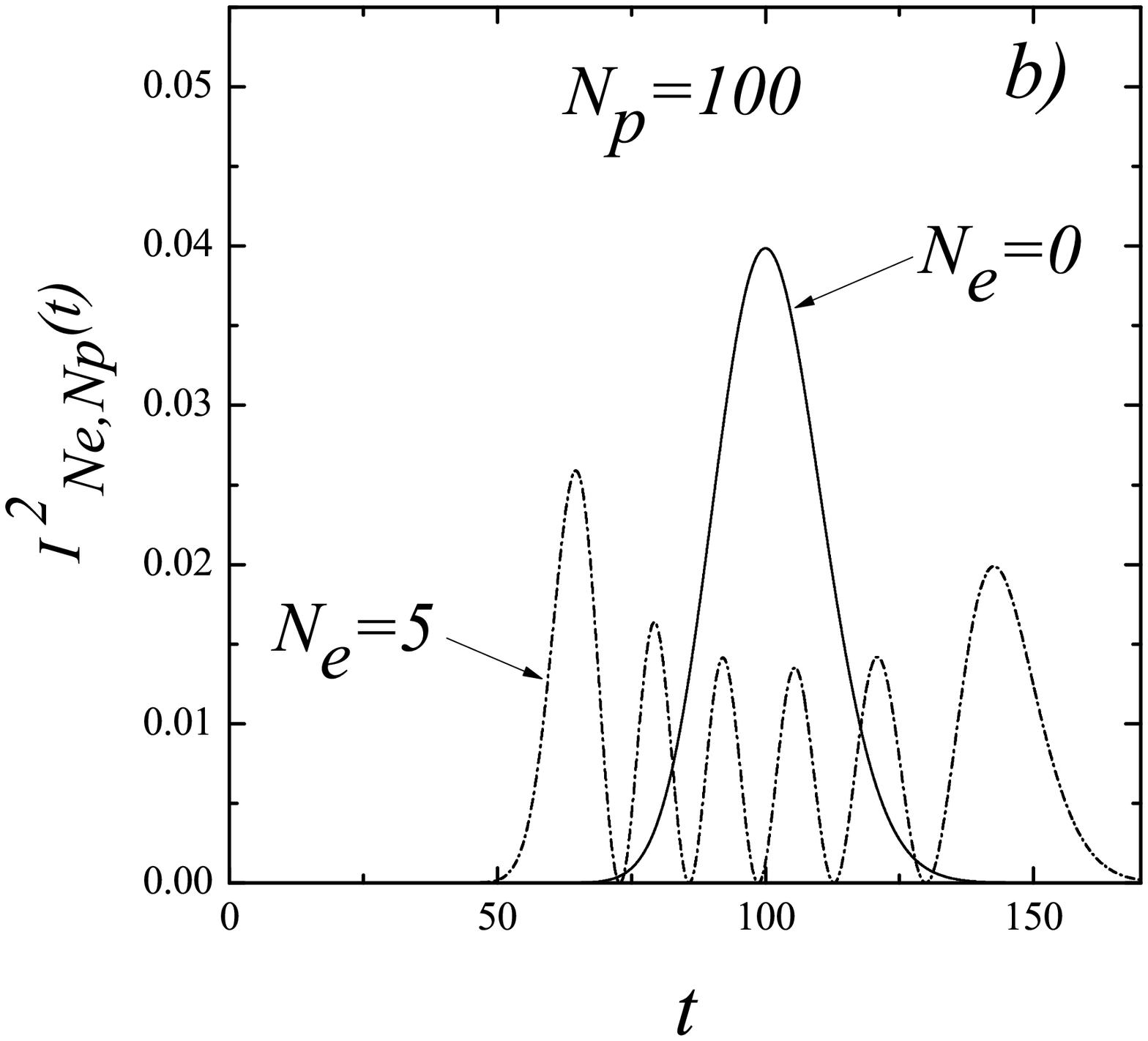}
\vskip -5mm
\caption{Some values for the $I^{2}_{N_e,N_p}(t)$--function as defined in Eq.~(\ref{funcI}).}
\label{figme5}
\end{center}
\end{figure}

We have considered the absorption structure function with some detail, because it helps us to understand the mean free path. Another ingredient is the function $I^{2}_{N_e,N_p}(t)$ (see Eqs.~(\ref{funcI})), which we have plot in Fig.~\ref{figme5} for different values of $N_p$ and $N_e$. As discussed in Section~\ref{nacs}, this function is part of the wave function of charged particles in a constant magnetic field: the energy levels are quantized for an axis perpendicular to the magnetic field direction and has continuum values parallel to the field. This is a function of $t=\omega^{2}_{\perp}/2 e B$ and in the panel $a)$ in this figure, we consider different values for $N_p$ with a fixed $N_e=0$. In panel $b)$ we take $N_p=100$ for two values of $N_e$. Our concern is how this function affects the result for the neutrino mean free path. Keeping in mind that $\int^{\infty}_{0} \, dt \, I^{2}_{N_e,N_p}(t) = 1$, the weight of this function is linked to the maximum value for $t$. The maximum value for $\omega_{\perp}=[(p_{\, n,x}+p_{\nu, x})^{2}+(p_{\, n,y}+p_{\nu, y})^{2}]^{1/2}$, results from the particle distribution function $f_{s_n}(E_{n},\mu_n, T)$ and the neutrino momentum. For the same $\omega^{max}_{\perp}$, different values for the magnetic field give different $t^{max}$. Together with the structure function, this $t^{max}$--value establish a constrain over the maximum values for $N_p$ and $N_e$.

We turn now to the analysis of the neutrino absorption mean free path. We conclude our study by adding the dispersion contribution, which have been discussed in~\cite{To19}. The behavior of these two contributions with temperature and with the magnetic field is very different. Due to this and for the benefit of the reader, we recall some aspects of the dispersion cross section in the following paragraphs. The presence of a constant magnetic field, establishes a preferred direction in space and consequently, the total cross section depends both on the magnitude of the momentum of the incoming neutrino and on the angle $\theta_{\nu}$ between its momentum and the direction of the magnetic field. For the dispersion reaction, an incoming angle of $\theta_{\nu}=\pi/2$ results in a cross section almost identical to the one in the absence of the magnetic field. This is because the phase space for this reaction is barely modified by the magnetic field. As we show soon, this is not the case for the absorption reaction, where the phase space (of final states) is substantially modified by the magnetic field.

In first place, in Eq.~(\ref{cs2}) we sum over all spin components. However and by taking for simplicity the $N_e=0$ case, the weak dynamics from Eq.~(\ref{trazaNe0}) already gives us some relevant information about this sum. In Table~\ref{asymmt}, we show results from Eq.~(\ref{trazaNe0}), where we have used $g_V=0.973$ and $g_A=1.197$. From this table, we can see that contributions with the spin down for the proton are zero for $du$ and are almost negligible for $dd$. Moreover, for the two extreme values for $\theta_{\nu}$, only one spin component contributes to the cross section: the $uu$--component for $\theta_{\nu}=0$ and the $ud$--component for $\theta_{\nu}=\pi$. Each component is weighed by a different factor, even thought these factors are similar in magnitude. This fact, together with the different shapes for the spin components of the absorption structure function shown in Fig.~\ref{figme2}, contribute to the asymmetry in the neutrino absorption cross section. Another ingredient is the partial polarization of the system, which is represented by the spin asymmetry ${\it A}$.
\begin{table}[h]
\begin{center}
\caption{Some values for the function $[L_{\mu\nu} N^{\mu\nu}/I^2_{0,N_p }(t)](s_p,s_n,\cos(\theta_{\nu}))$ from Eq.~(\ref{trazaNe0}) for $p_{e, \,z}>0$. Note that this function has no dimensions.}
\label{asymmt}
\vskip 2mm
\begin{tabular}{ccccc}   \hline\hline
~~~$s_p, \, s_n$~~~  &~~~~$\theta_{\nu}=0$~~~~&~~~~$\theta_{\nu}=\pi/2$~~~~&~~~~$\theta_{\nu}=\pi$~~~~&  \\ \hline
$uu$       &  $18.84$  &$9.42$  & $~0.$  &              \\
$ud$       &  $0.$     &$11.46$  & $~~22.92$  &              \\
$du$       &  $0.$     &$0.$  & $\sim 0.$  &              \\
$dd$       &  $~~0.20$   &$~~0.10$  & $~~~0.$  &              \\
\hline\hline
\end{tabular}
\end{center}
\end{table}

\begin{figure}[t]
\begin{center}
\vskip -10mm
    \includegraphics[scale=0.47]{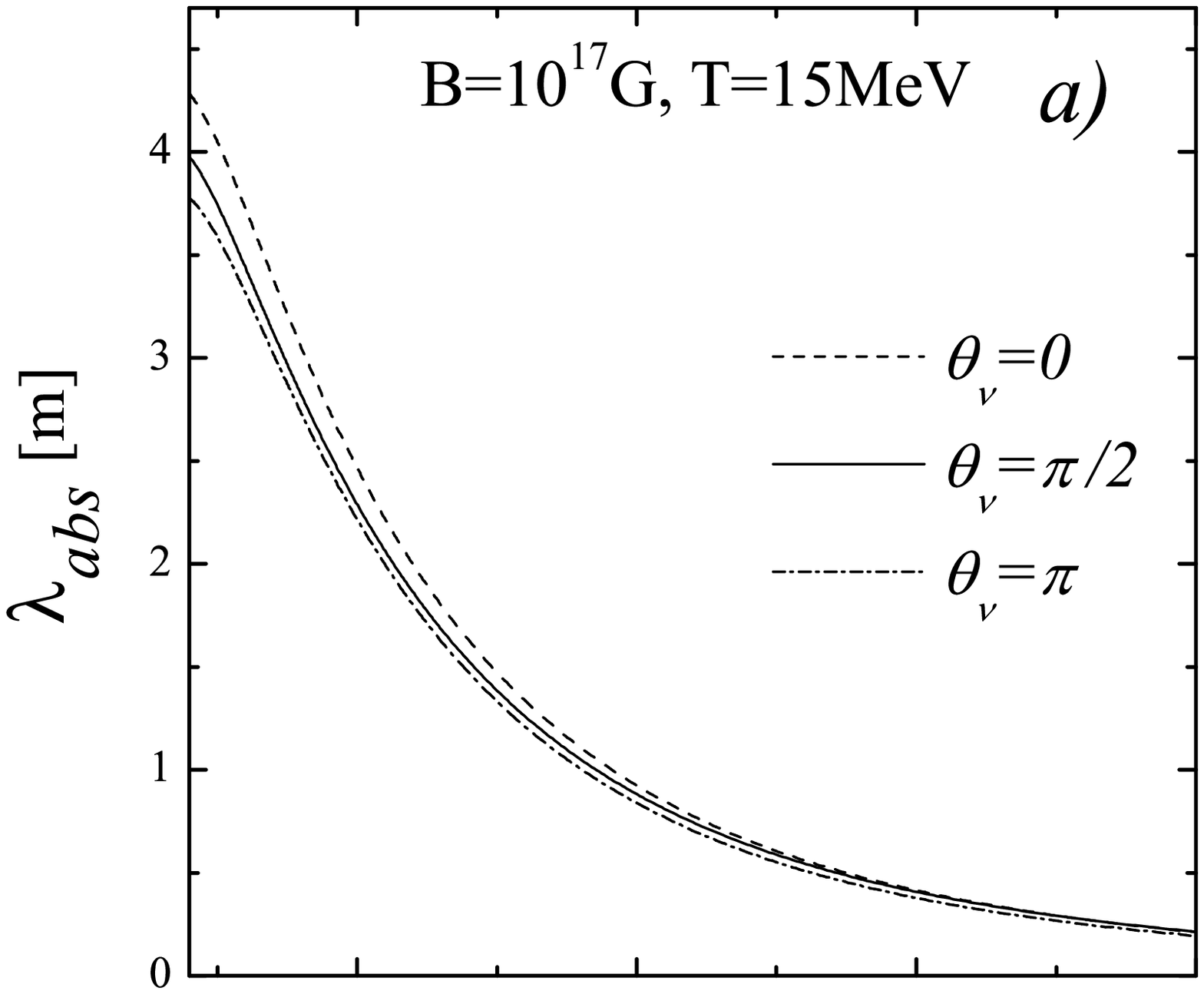}
\vskip -6.1cm
    \includegraphics[scale=0.47]{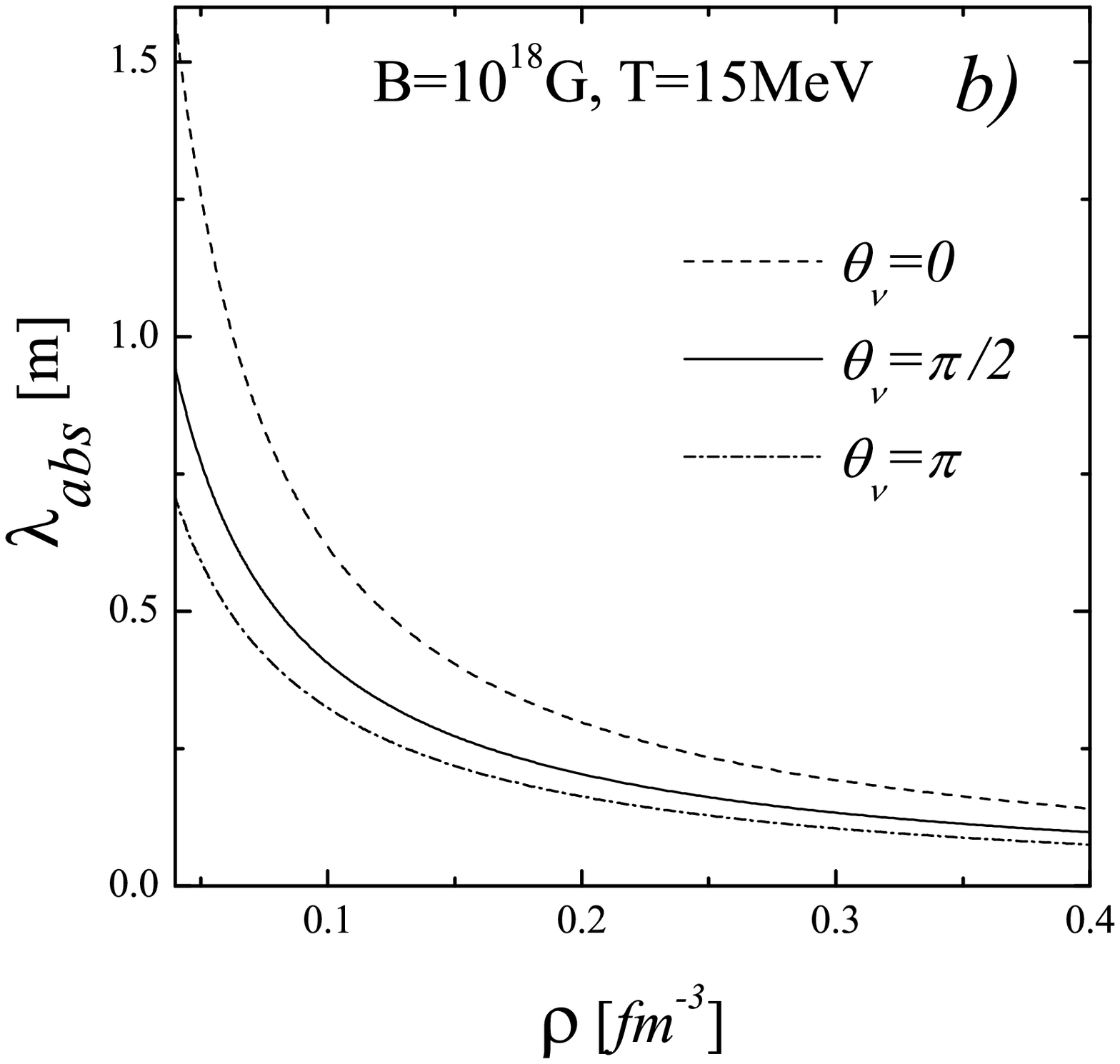}
\vskip -7mm
\caption{The absorbtion neutrino mean free path as a function of the density and for three different values for the neutrino incoming angle, $\theta_{\nu}$. In panel $a)$ we show results for a magnetic field intensity $B=10^{17}$G, while we have $B=10^{18}$G for panel $b)$. The momentum of the incoming neutrino is $|\vec p_{\nu}| = 3 T$.}
\label{figme6}
\end{center}
\end{figure}

In Fig.~\ref{figme6}, we present our result for the absorption neutrino mean free path as a function of the density, at a temperature T$=15$MeV, for two values of the magnetic field B$=10^{17}$G and B$=10^{18}$G and for three different angles of the incoming neutrino. If we compare these results with the dispersions ones (see Fig.~10 in~\cite{To19}), we notice that the mean free path shows the same qualitative behavior. But, at variance with the dispersion case, the magnitude of the absorption mean free path has a strong dependence with the magnetic field. From B$=10^{17}$G to B$=10^{18}$G there is an important reduction in the mean free path. The reason for this reduction is due to the magnetic dependence of the phase space for final states. An increase of this phase space result in an increase of the cross section and consequently a reduction in the mean free path. As we have already discussed, when the magnetic field grows, the number of the Landau levels which contribute to the cross section decrease. But the degeneracy of the levels, given by $e B A/2 \pi$, grows. Therefore, for increasing values of the magnetic field there is some kind of competition between the increase of the final phase space due to the degeneracy and the reduction in the number of Landau levels. From our numerical results, it turns out that within a range for the magnetic field between B$=10^{16}$G up to B$=10^{18}$G, the absorption neutrino mean free path decrease for increasing values of the magnetic field. Referring now to the maximum values for $N_p$ and $N_e$, we can give only indicative values, as they change with density (they depend also on the temperature, on the single particle energies and on the chemical potential). For $\rho=0.16$fm$^{-3}$, we have $N_p \simeq 150$ and $N_e \simeq 10$ for B$=10^{17}$G, while the values for B$=10^{18}$G are $N_p \simeq 15$ and $N_e = 0$.

\begin{figure}[t]
\begin{center}
\vskip -10mm
    \includegraphics[scale=0.47]{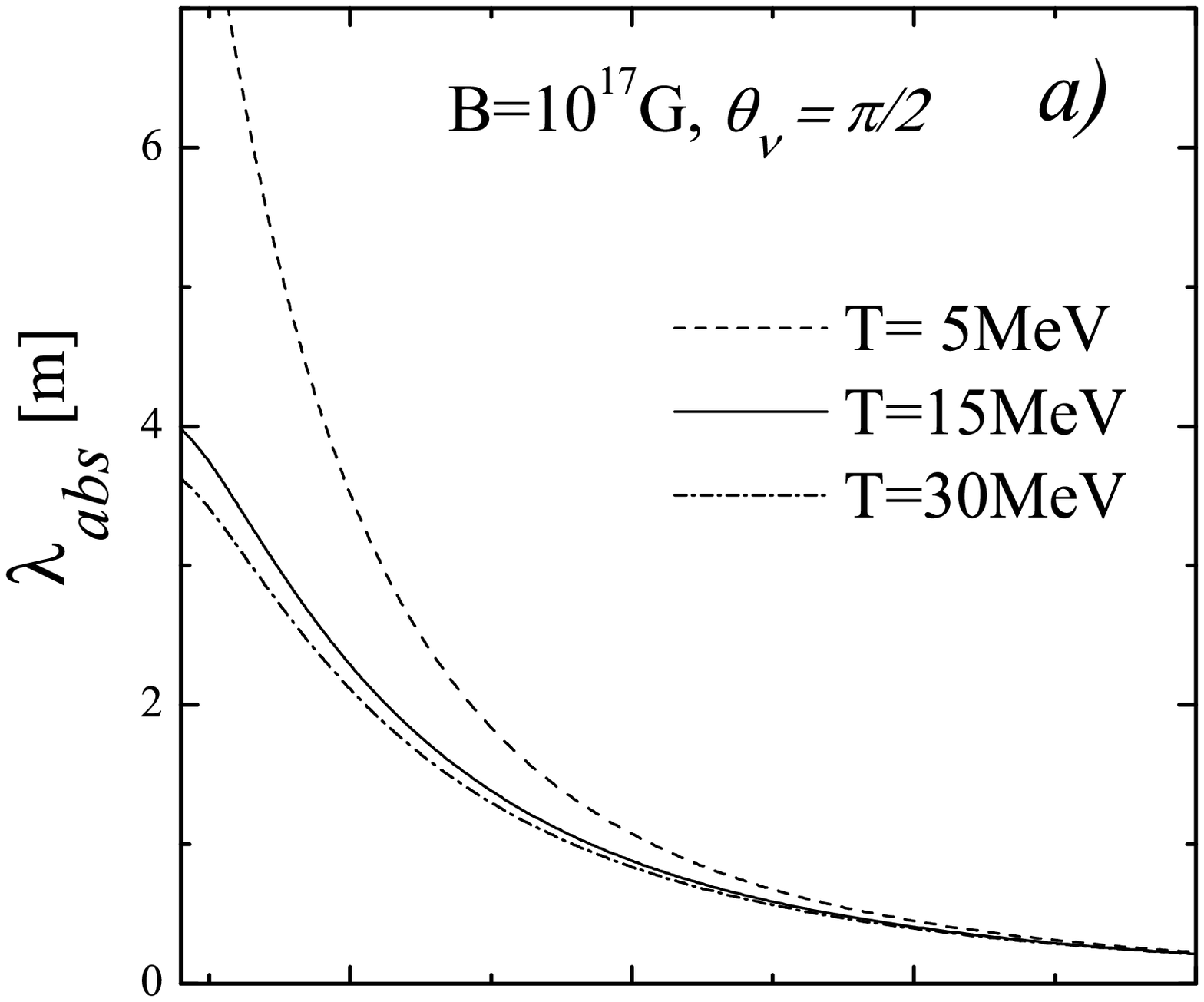}
\vskip -6cm
    \includegraphics[scale=0.47]{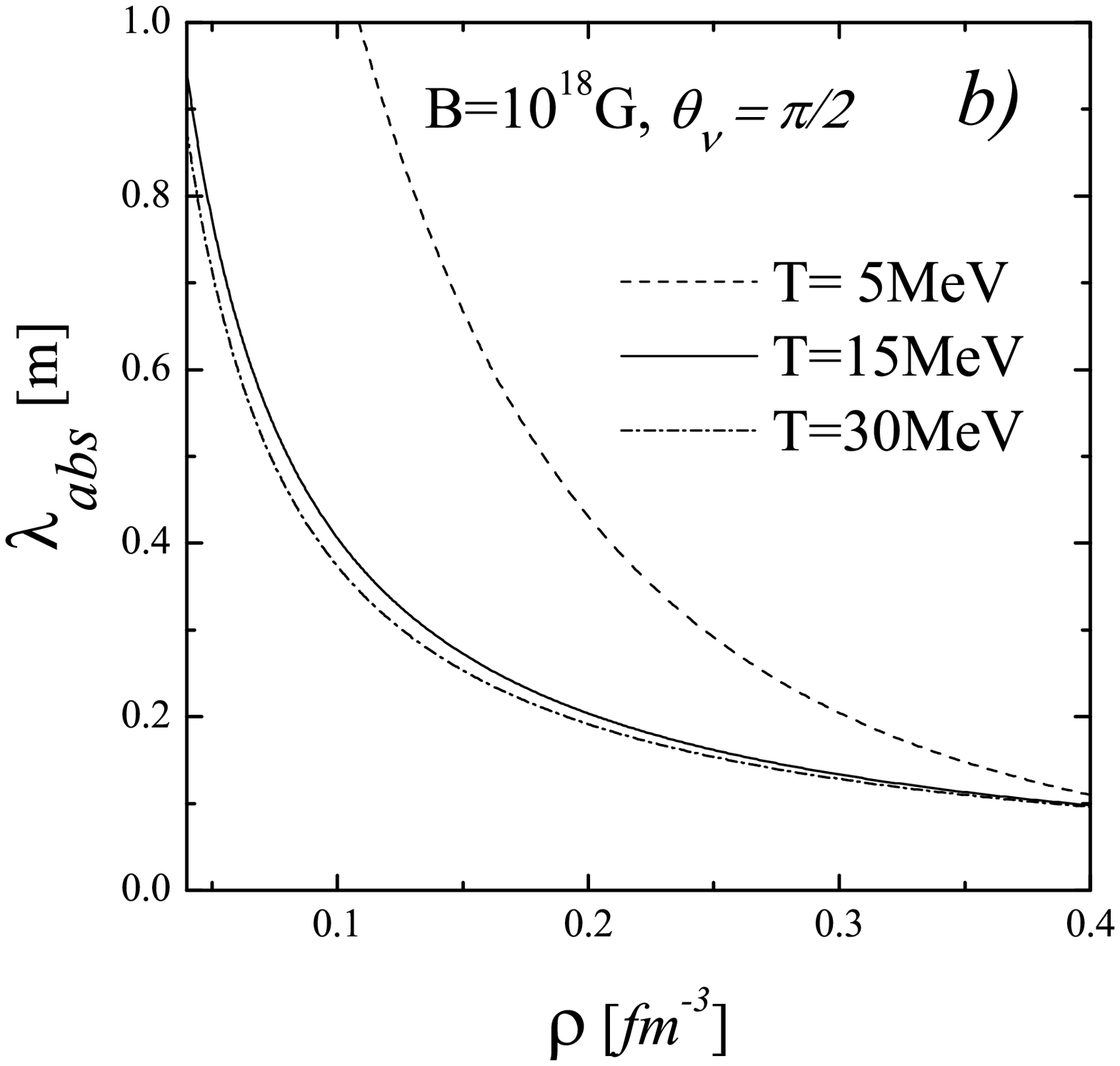}
\vskip 2mm
\caption{The absorbtion neutrino mean free path for three different values for the temperature. As in Fig.~~\ref{figme6}, the panel $a)$ ($b$) is the results for a magnetic field intensity $B=10^{17}$ ($B=10^{18}$G), using the same approximation for the momentum of the incoming neutrino.}
\label{figme7}
\end{center}
\end{figure}

In the next step, we analyze the temperature dependence of the absorption neutrino mean free path. In Fig.~\ref{figme7}, we consider three temperatures: T$=5, \; 15$ and $30$MeV, for B$=10^{17}$G and B$=10^{18}$G. For simplicity, we have plotted only the results for $\theta_{\nu}=\pi/2$ and for the energy of the neutrino we have used the prescription $|\vec p_{\nu}| = 3 T$. Our results show that the temperature dependence is rather weak, specially when compared with the dispersion case. To understand this behavior it is useful to compare the dispersion structure function in Fig.~3,~\cite{To19} with the absorption ones in Fig.~\ref{figme4}: the area of the absorption  structure function decreases, instead of increasing. This means that the absorption mean free path should increase for higher temperature values. However, our structure function spread over a wider energy region as the temperature grows, populating more Landau levels. The increase in the number of Landau levels turn down the value of the mean free path. The combined result is a small decrease in the absorption mean free path with temperature.

\begin{figure}[t]
\begin{center}
    \includegraphics[scale=0.53]{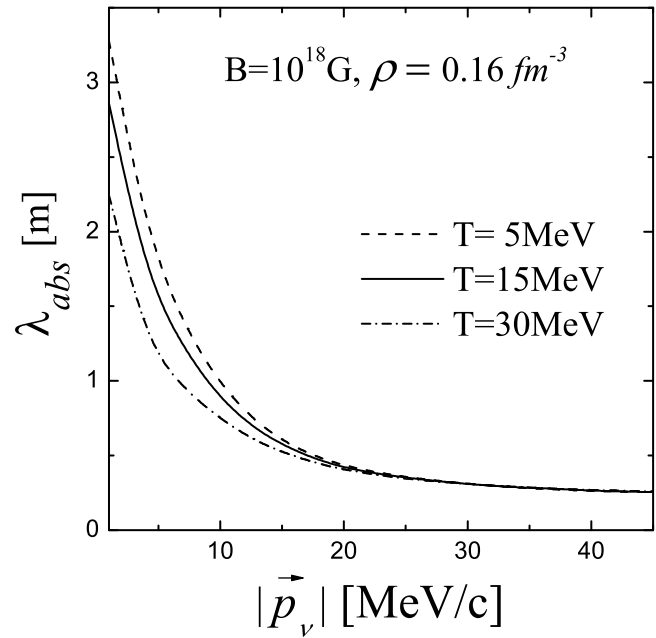}
\vskip 2mm
\caption{The absorbtion neutrino mean free path as a function of the momentum of the incoming neutrino $|\vec p_{\nu}|$, for $\theta_{\nu}=\pi/2$. We have chosen three values for the temperature.}
\label{figme8}
\end{center}
\end{figure}

The temperature dependence is further explored in Fig.~\ref{figme8}, where the neutrino absorption mean free path is depicted as a function of the momentum of the neutrino for three values of the temperature, B$=10^{18}$G, a density $\rho=0.16$fm$^{-3}$ and $\theta_{\nu}=\pi/2$. The $|\vec p_{\nu}|$--dependence of the neutrino mean free path shows a qualitative agreement for both the dispersion and absorption reactions. This is because the structure function is larger for larger values of the momentum of the neutrino. For the absorption reaction, for an increasing value for $|\vec p_{\nu}|$ we have more energy in the initial state and therefore more Landau levels contribute to the mean free path. The reduction in the structure function for higher temperatures obviously remains. The interplay among these elements for the absorption reaction, results in a neutrino mean free path almost independent of the temperature. This is a particular result and we can not give a deeper explanation. Having in mind the rule $|\vec p_{\nu}| = 3 T$ and going back to Fig.~\ref{figme7}, we notice the same result: the mean free path for T$=5$MeV ($|\vec p_{\nu}|=15$MeV) is clearly separated from the ones for T$=15$ ($|\vec p_{\nu}|=45$MeV) and $30$MeV ($|\vec p_{\nu}|=90$MeV), for all densities.

\begin{figure}[t]
\begin{center}
\vskip -30mm
    \includegraphics[scale=0.47]{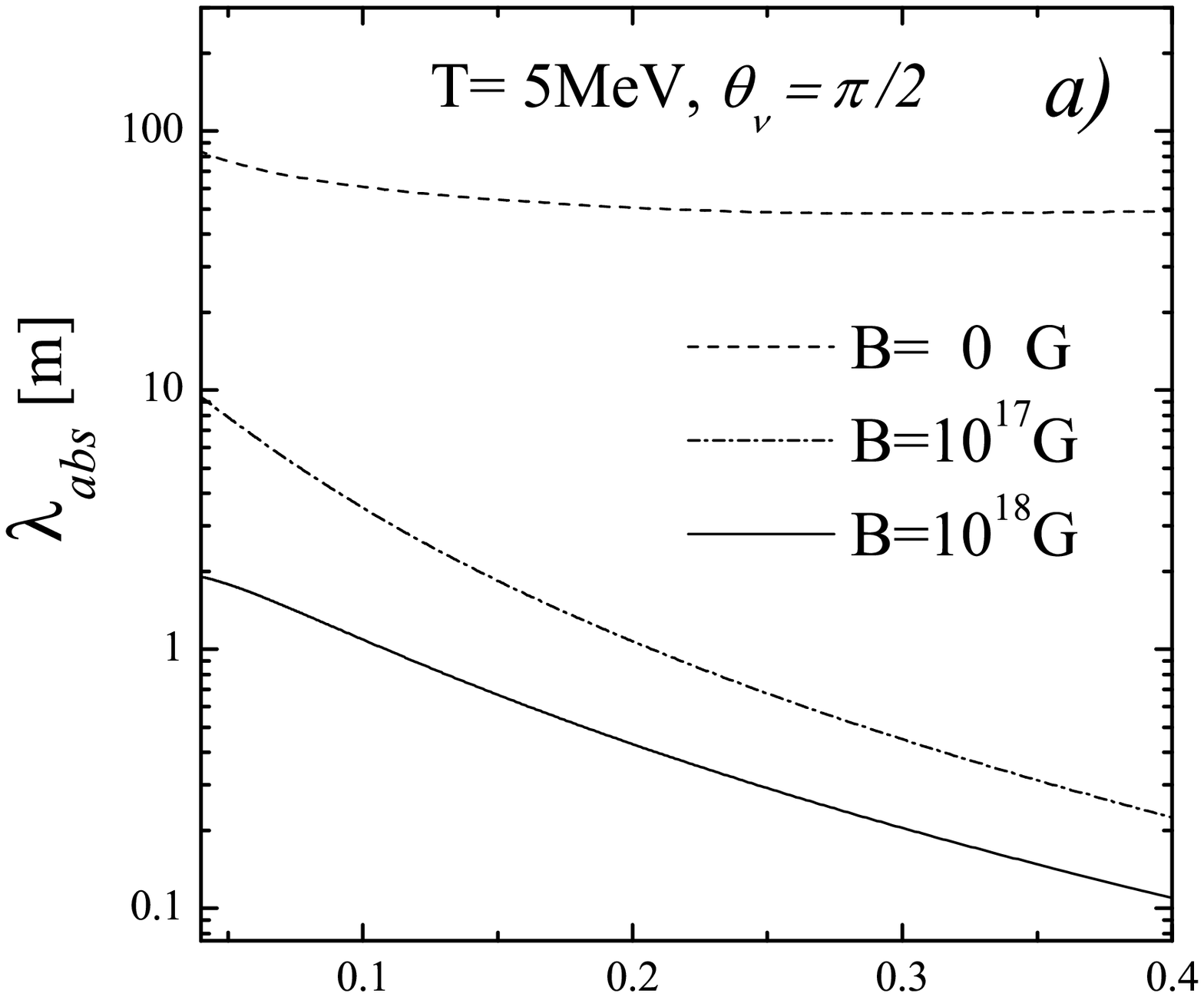}
\vskip -6cm
    \includegraphics[scale=0.47]{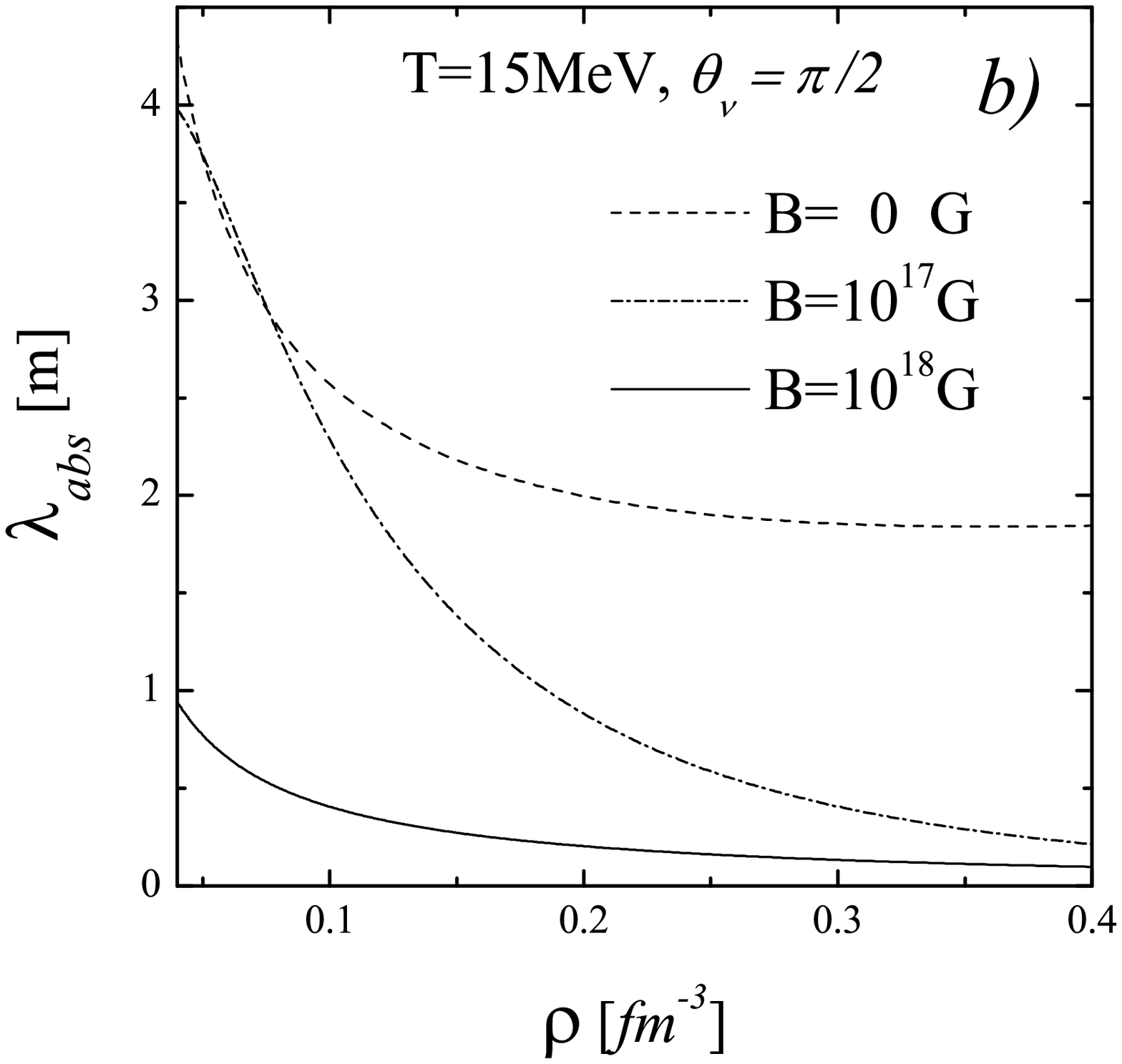}
\vskip -5mm
\caption{The absorbtion neutrino mean free path as a function of the density and for three different values for the magnetic field intensity. For the neutrino incoming angle we have employed $\theta_{\nu}=\pi/2$ and we have used $|\vec p_{\nu}| = 3 T$. In panel $a)$ the temperature is T$=5$MeV, while in panel $b)$ we have T$=15$MeV.}
\label{figme9}
\end{center}
\end{figure}

At this point, it is clear that the phase space for the final state in the absorption reaction is very different from the one in the dispersion reaction due to the magnetic field. The magnetic field can be reduced continuously up to B$=0$. In the absence of magnetic field, the phase space for absorption and for the dispersion reaction is the same~\cite{ReddyT}. In Fig.~\ref{figme9}, we show the absorption neutrino mean free path for magnetic fields B=$0$, $10^{17}$G and $10^{18}$G, $\theta_{\nu}=\pi/2$ and two temperatures: T=$5$MeV in panel $a)$ and T=$15$MeV in panel $b)$. Note that in panel $a)$ we have employed a logarithmic scale for $\lambda_{abs}$. The absorption mean free path for B$=0$ has a different functional dependence with the density and a very pronounced temperature--dependence, consistent with the one for the dispersion reaction. Let us recall that the phase space for the dispersion reaction is barely affected by the magnetic field. It is not a trivial subject to perform the limit from a strong magnetic field to B$=0$. This discussion goes beyond the scope of the present contribution and we refer the reader to~\cite{Ar99,Sh05} for details on how to perform this limit process.

\begin{figure}[t]
\begin{center}
\vskip -30mm
    \includegraphics[scale=0.47]{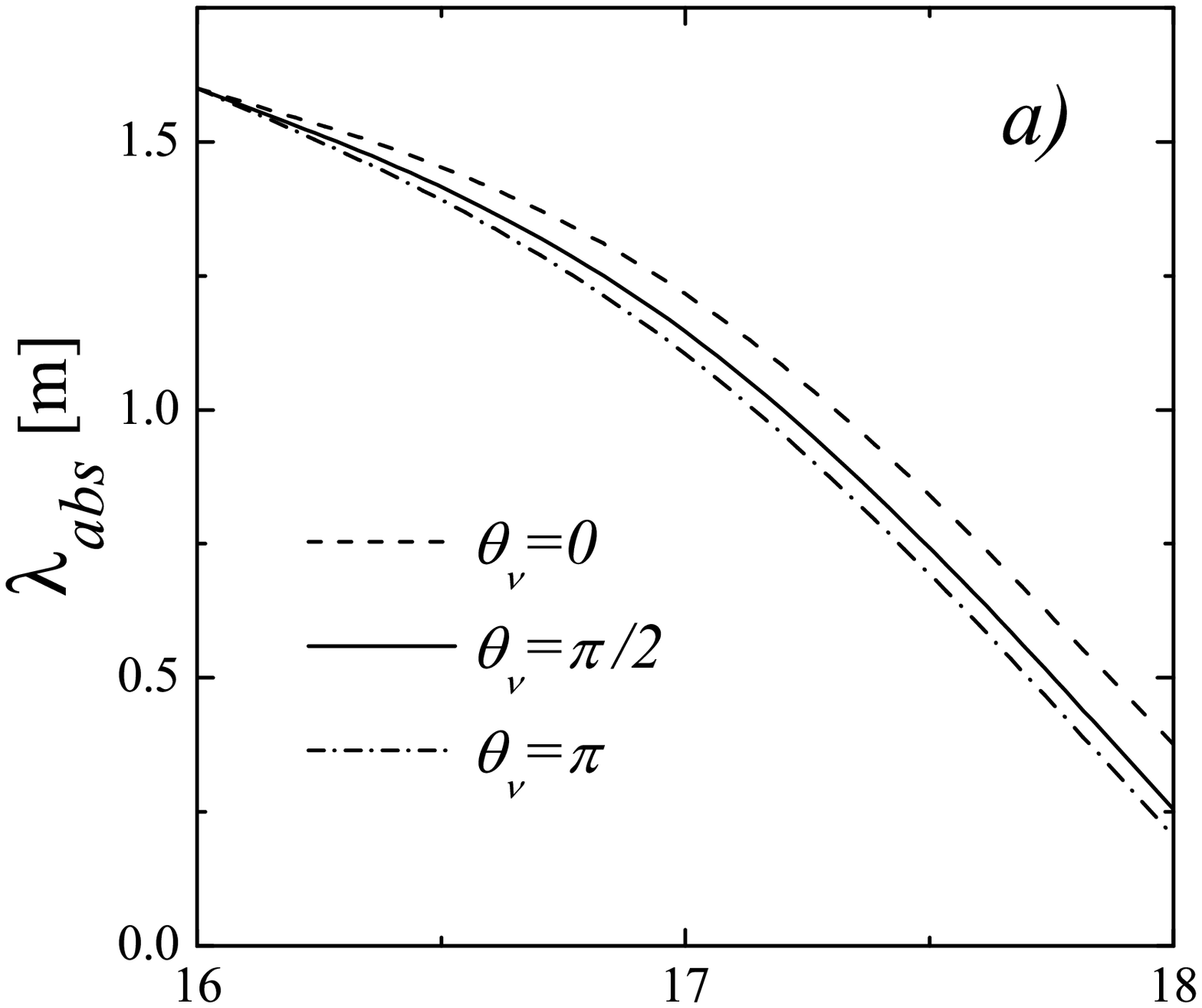}
\vskip -6cm
    \includegraphics[scale=0.47]{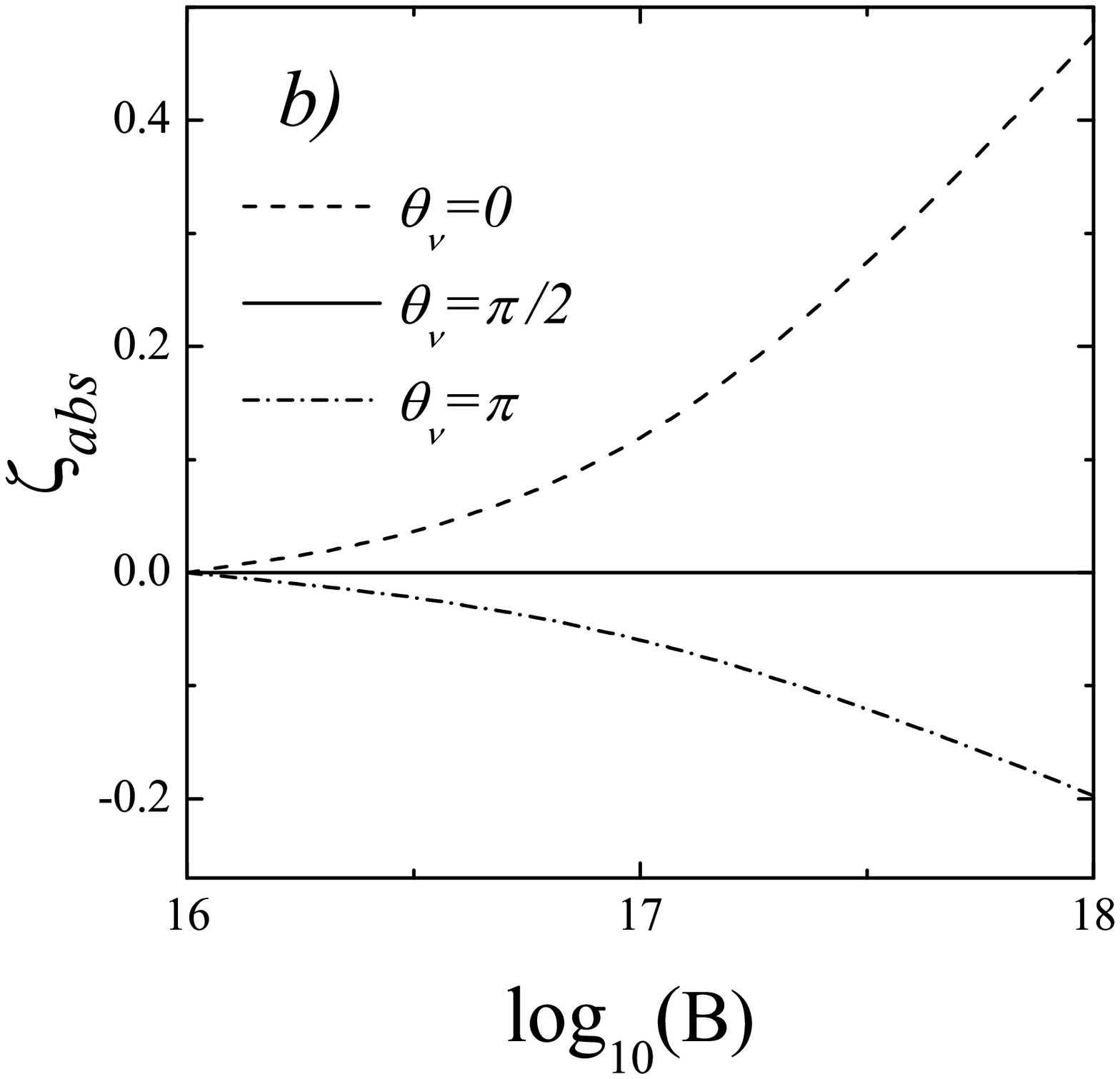}
\vskip -5mm
\caption{Dependence on the magnetic field intensity. We have fixed the density at $\rho=0.16$ $fm^{-3}$ and T$=15$~MeV. The absorption mean free path is depicted in panel $a)$, for three angles the incoming neutrino $\theta_\nu$, while in panel $b)$ we show $\zeta_{abs}$ as defined in Eq.~(\ref{mfpasym}), for the same set of angles. Units of the magnetic field intensity B, is given in Gauss.}
\label{figme10}
\end{center}
\end{figure}

In what follows, we focus on the asymmetry of the neutrino mean free path. In the panel $a)$ in Fig.~\ref{figme10}, we show $\lambda_{abs}$ as a function of the magnetic field intensity. This is done at a density $\rho=0.16$fm$^{-3}$, T$=15$MeV and for three angles: $\theta_{\nu}=0$, $\pi/2$ and $\pi$. As the magnitude of $\lambda_{abs}$ decreases for increasing values of the magnetic field, this figure is somehow misleading because the asymmetry is not clearly seen. Due to this, we have defined the quantity,
\begin{equation}
\zeta_{abs} =\frac{\lambda_{abs}(\theta_{\nu})-\lambda_{abs}(\theta_{\nu}=\pi/2)}{\lambda_{abs}(\theta_{\nu}=\pi/2)} \ ,
\label{mfpasym}
\end{equation}
which gives a more accurate idea of the increase of the asymmetry in the neutrino mean free path. The $\zeta_{abs}$--function is depicted in the panel $b)$ in the same figure. As already discussed, the magnetic field establish a preference axis in space. Our results show that it is more likely for a neutrino moving antiparallel to the magnetic field ($\theta_{\nu}=\pi$) to be absorbed, than a one which moves parallel to it. Assuming an isotropic production of neutrinos, this implies that more neutrinos are emitted parallel to the magnetic field. In an actual neutron star model, the whole magnetic field can not be considered as a constant vector field. Our model should be applied locally, according to the geometry of the field.

\begin{figure}[t]
\begin{center}
    \includegraphics[scale=0.53]{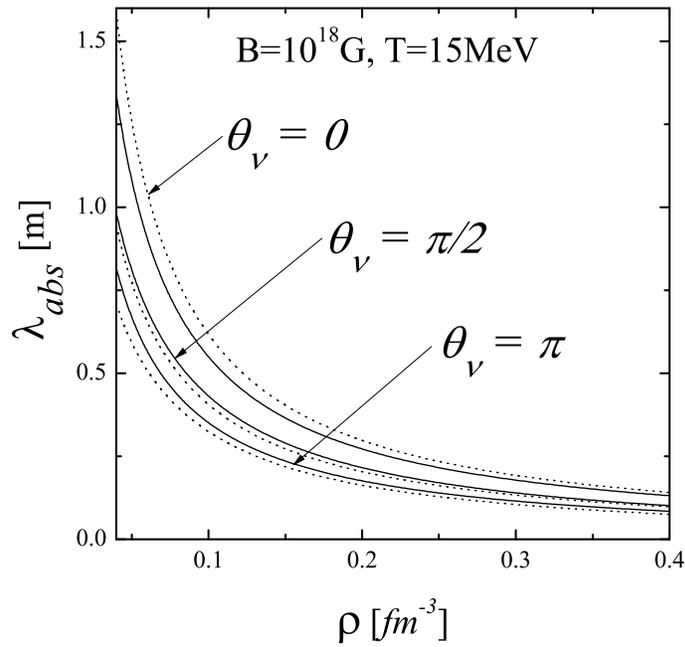}
\vskip 2mm
\caption{The absorbtion neutrino mean free path as a function of the density and for three different values for the neutrino incoming angle, $\theta_{\nu}$, $B=10^{18}$G
and T$=15$MeV, where the momentum of the incoming neutrino is taken as $|\vec p_{\nu}| = 3 T$. The continuous lines are the case where the spin asymmetry ${\it A}$, is arbitrarily taken as zero, while for the dotted lines we employed the not--null {\it A}--value from our EoS.}
\label{figme11}
\end{center}
\end{figure}

The asymmetry in the mean free path for both the absorption and for the dispersion reactions, results from the interplay among several elements. Considering the different interactions which take place in the process, we have: $i)$ the results from Table~\ref{asymmt}, give us information on the weak--interaction contribution to the asymmetry in the mean free path. $ii)$ the strong--interaction, which favors the situation ${\it A}=0$ and $iii)$ the coupling of the magnetic field with protons, neutrons and electrons, which tends to polarized the system. The balance among these two last elements leads to the equilibrium values for the spin asymmetry ${\it A}$, the effective masses and the chemical potential. For simplicity, sometimes all these contributions are summarized in one single quantity: the spin asymmetry ${\it A}$. In Fig.~\ref{figme11}, we show the mean free path, under the same conditions of panel $b)$ in Fig.~\ref{figme6}, but evaluating the neutrino mean free path putting arbitrarily ${\it A}=0$ (continuous lines in the figure). For comparison we give also the results from Fig.~\ref{figme6} (dotted lines). We can see that the isolated contribution from ${\it A}$, does not explain the main contribution to the mean free path asymmetry. Our point here, is that the evaluation of the asymmetry in the neutrino mean free path requires a consistent model, starting from the EoS and considering all the just mentioned elements.

\begin{figure}[t]
\begin{center}
\vskip -10mm
    \includegraphics[scale=0.47]{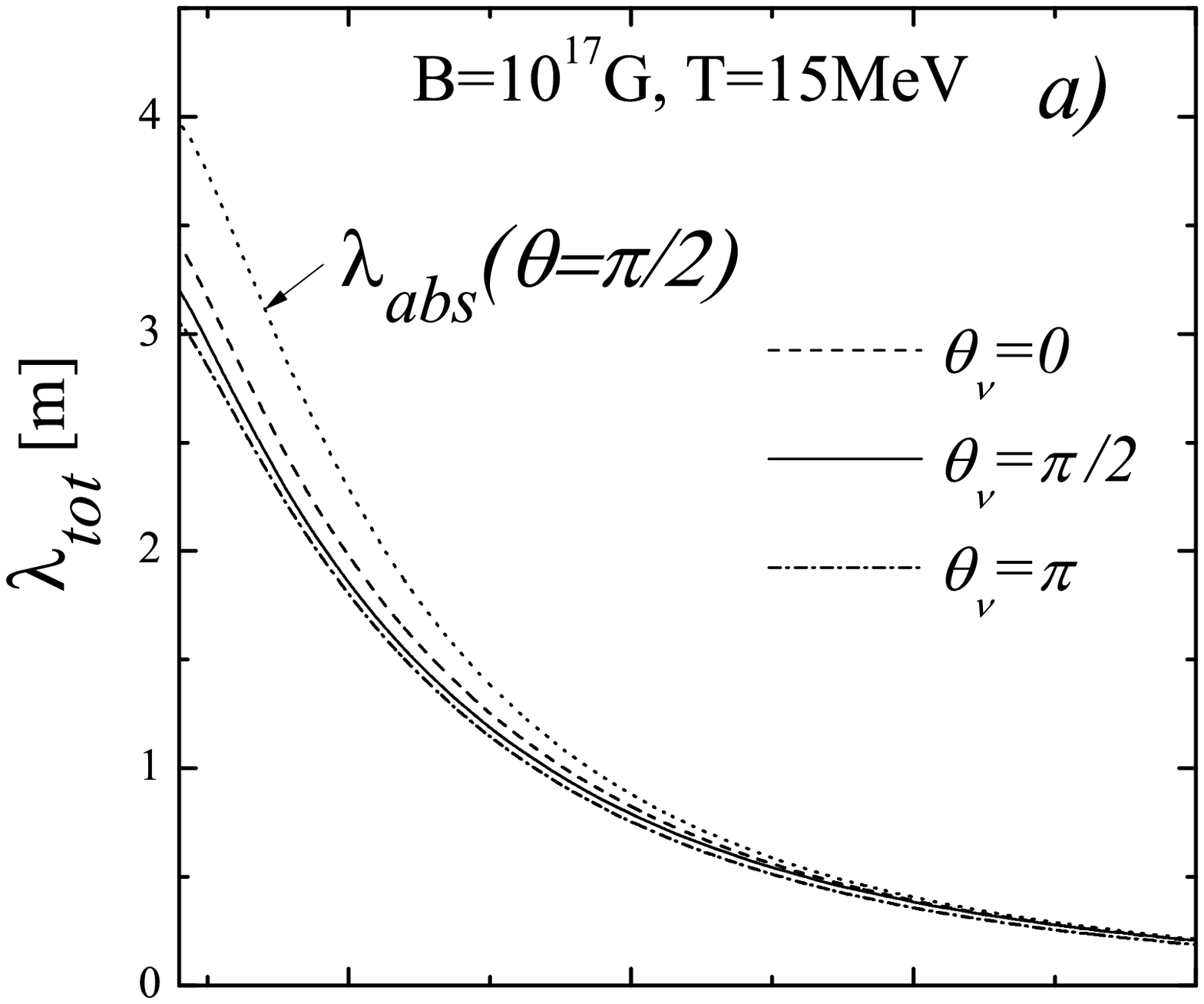}
\vskip -6cm
    \includegraphics[scale=0.47]{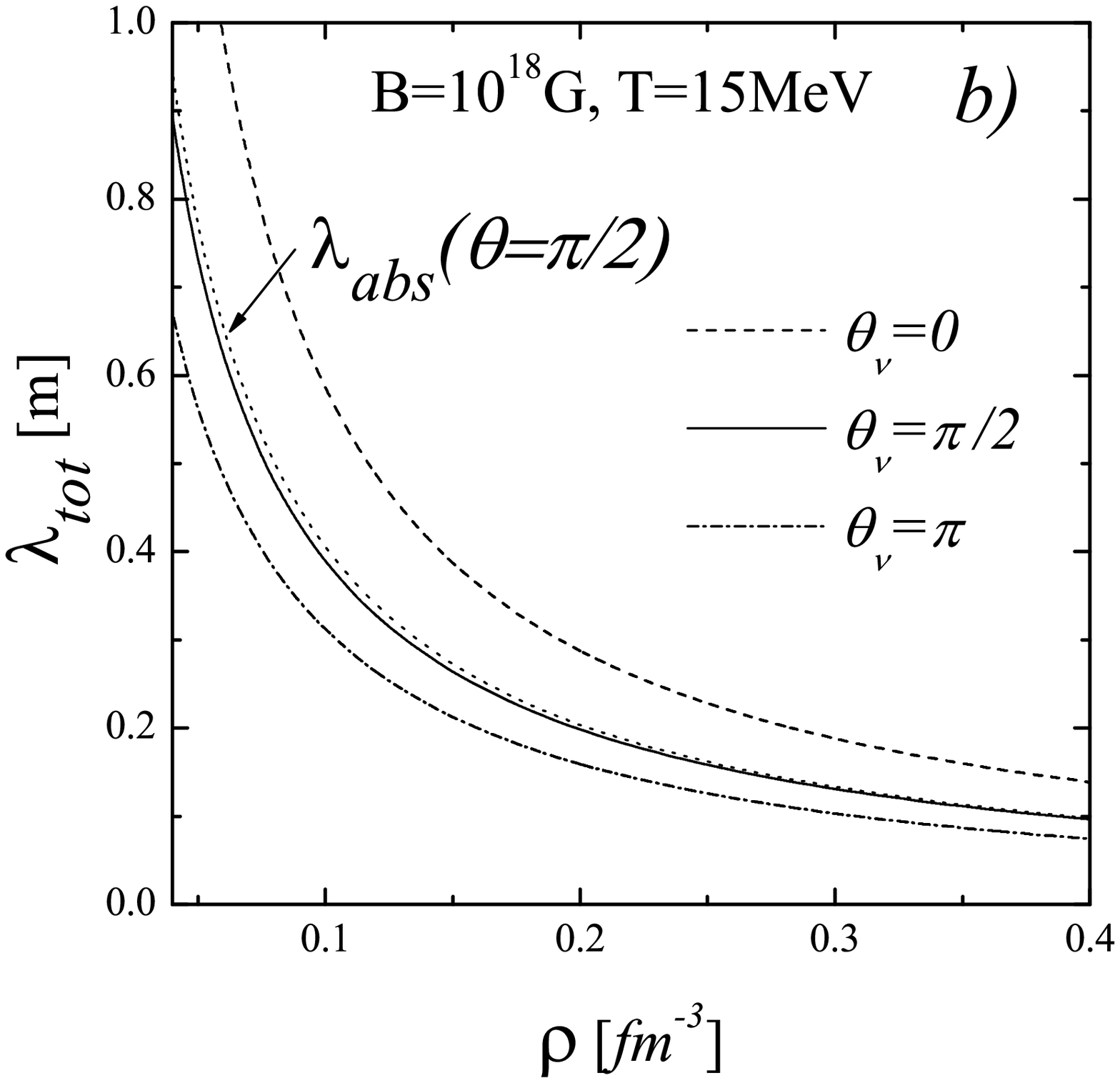}
\vskip 2mm
\caption{The total neutrino mean free path for three different values for the neutrino incoming angle, $\theta_{\nu}$ and for T$=15$MeV. As in Fig.~\ref{figme6}, the panel $a)$ ($b$) is the results for a magnetic field intensity $B=10^{17}$ ($B=10^{18}$G), using the same approximation for the momentum of the incoming neutrino. For convenience, we show also the absorption neutrino mean free path for $\theta_{\nu}=\pi/2$.}
\label{figme12}
\end{center}
\end{figure}

\begin{figure}[t]
\begin{center}
\vskip -10mm
    \includegraphics[scale=0.47]{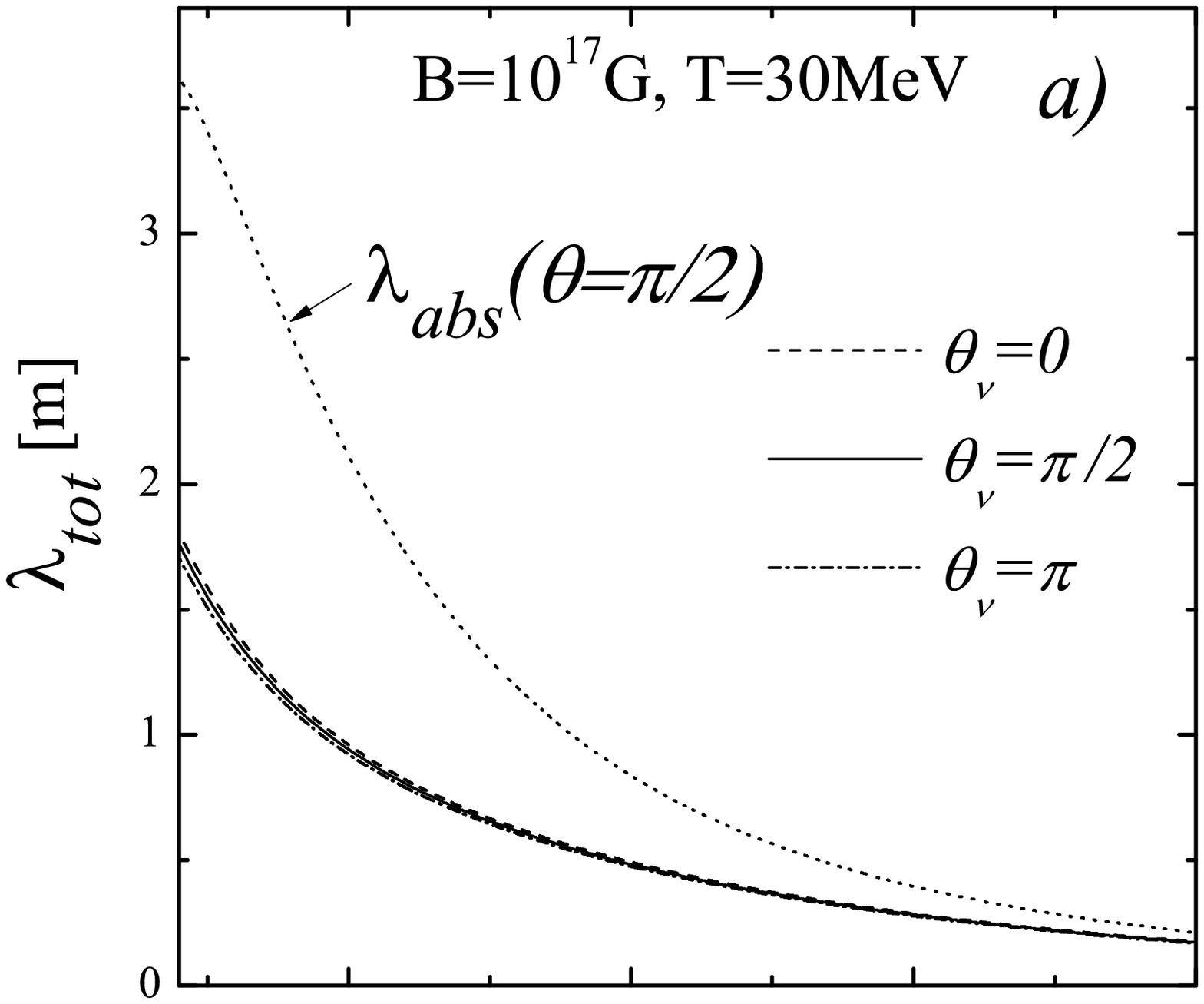}
\vskip -6cm
    \includegraphics[scale=0.47]{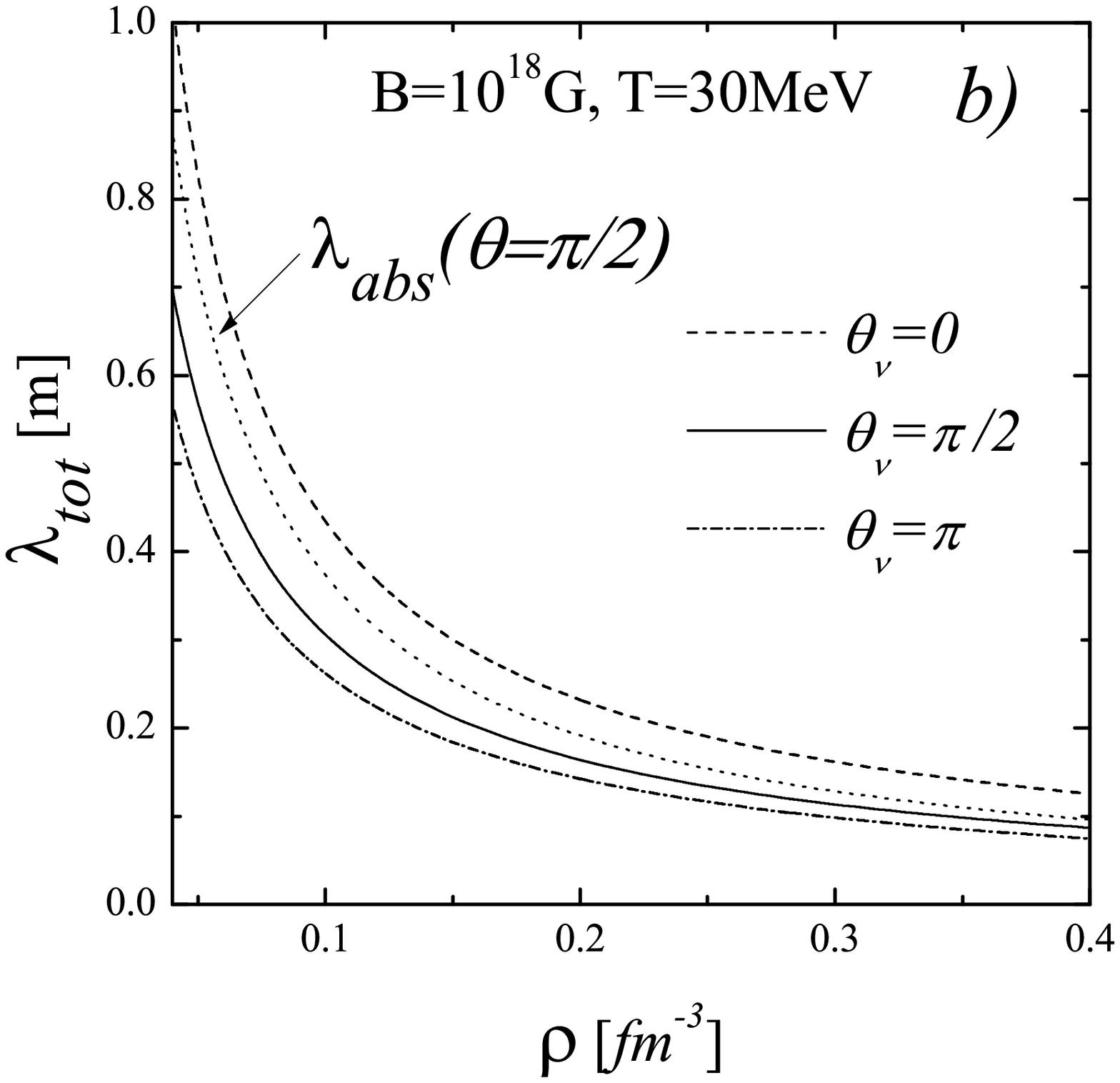}
\vskip 2mm
\caption{The same as in Fig.~\ref{figme12}, but for T$=30$MeV.}
\label{figme13}
\end{center}
\end{figure}

As a final point, we include the dispersion contribution to the mean free path. The addition of this contribution gives the total neutrino mean free path, $\lambda_{tot}$,
\begin{equation}
\lambda_{tot} = \left(\frac{1}{\lambda_{abs}}+\frac{1}{\lambda_{dis}} \right)^{-1}.
\label{mfptot}
\end{equation}
Results for $\lambda_{dis}$ have been taken from~\cite{To19}.
We give our results for this quantity in Figs.~\ref{figme12} and \ref{figme13}. In the first figure we show $\lambda_{tot}$ as a function of the density, for B$=10^{17}$ and $=10^{18}$G, three angles for the incoming neutrino: $\theta_{\nu}=0$, $\pi/2$ and $\pi$ and a temperature T$=15$MeV. The second figure has the same variables except for the temperature where we have employed T$=30$MeV. In both figures we have included also $\lambda_{abs}$ for $\theta_{\nu}=\pi/2$. This is done as a reference of the relative importance of the absorption contribution. Before we go on with our analysis, it is worth to recall that $\lambda_{abs}$ and $\lambda_{dis}$ have very different behavior for the temperatures and the magnetic fields considered in the present contribution. While $\lambda_{dis}$ has a strong dependence with temperature and its value for $\theta_{\nu}=\pi/2$ is almost independent of the magnetic field, $\lambda_{abs}$ has a weak dependence with temperature and it decreases for increasing values of the magnetic field. This contrasts with the result for B$=0$: in this case both $\lambda_{abs}$ and $\lambda_{dis}$ have the same (strong) dependence with temperature and due to the values of the coupling constants, one has $\lambda_{abs} < \lambda_{dis}$.

By comparing now the panel $a)$ and $b)$ in Fig.~\ref{figme12}, we notice that the dispersion reaction is as important as the absorption one for B$=10^{17}$G, while it is negligible for B$=10^{18}$G. This is because of the dependence of $\lambda_{abs}$ with the magnetic field. By doing the same comparison in Fig.~\ref{figme13}, we notice that the dispersion contribution becomes more important, due to the strong temperature dependence of $\lambda_{dis}$.

We want to finish the discussion on our results, by performing a quantitative analysis of the asymmetry. To this end, we define the mean free path asymmetry as,
\begin{equation}
\chi_{tot} =\frac{\lambda_{tot}(\theta_{\nu}=0)-\lambda_{tot}(\theta_{\nu}=\pi)}{<\lambda_{tot}(\theta_{\nu})>} \ ,
\label{chimfp}
\end{equation}
where we have employed $<\lambda_{tot}(\theta_{\nu})> \cong (\lambda_{tot}(\theta_{\nu}=0)+\lambda_{tot}(\theta_{\nu}=\pi))/2$. Note that for the dispersion reaction, one has $<\lambda_{tot}(\theta_{\nu})> = \lambda_{tot}(\theta_{\nu}=\pi/2)$. We give numerical values for $\chi_{tot}$ in Table~\ref{asymmtot}, for three values of the density and for B$=10^{17}$G and $10^{18}$G, with temperatures of T$=15$MeV and $30$MeV. As expected, the mean free path asymmetry is more important for the stronger magnetic fields. The reduction in $\chi_{tot}$ for higher values of the density is because the strong interaction becomes more important. Let us recall that the strong intereaction favors a non-polarized system. Some increase of $\chi_{tot}$ at $\rho=0.40$fm$^{-3}$ is particular to many of the Skyrme--parameterizations. Beyond this difficulty, we have preferred to employ the same parametrization as in~\cite{To19}, in order to make a fair comparison of both contributions to the total mean free path.

The increase of the temperature leads to a decrease in the mean free path asymmetry. This result seems intuitively correct, as temperature reduce the spin asymmetry ${\it A}$. However, it is convenient to give some details on the origin of this results. In first place, $\lambda_{abs}$ has a weak temperature--dependence. On the other hand, $\lambda_{dis}$ depends strongly with the temperature, but it mean free path asymmetry ($\chi_{dis}$), is rather independent of the temperature. The last element is that the absorption mean free path asymmetry is bigger than the dispersion one. This is because in the absorption reaction we deal with charged particles which have a stronger interaction with the magnetic field. Now, as temperature grows, the $\lambda_{dis}$ contribution to $\chi_{tot}$ becomes more important, which leads to smaller values for $\chi_{tot}$, which explains the temperature dependence of our results in this table.
\begin{table}[t]
\begin{center}
\caption{Mean free path asymmetry $\chi_{tot}$, as a function of the density for two values of the magnetic field intensity and two values of the temperature.}
\label{asymmtot}
\begin{tabular}{cccccccc}   \hline\hline
~~~$\rho$~[fm$^{-3}$]~~~  &~~~~& \mc {2}{c}{~~~~$\chi_{tot}(B=10^{17}G)$~~~~} &~~~~&\mc {2}{c}{~~~~$\chi_{tot}(B=10^{18}G)$~~~~} &~~~~
 \\ \cline{3-4}\cline{6-7}
              & & ~~~T$=15$MeV ~~~& ~~~T$=30$MeV~~~  &   & ~~~T$=15$MeV~~~ & ~~~T$=30$MeV~~~   &                \\ \hline
$0.050$       & & $0.112$&$0.068$  &   & $0.740$&$0.565$   &                \\
$0.160$       & & $0.088$&$0.034$  &   & $0.579$&$0.479$   &             \\
$0.400$       & & $0.094$&$0.042$  &   & $0.603$&$0.506$   &              \\
\hline\hline
\end{tabular}
\end{center}
\end{table}

In the last point for this section, we make some comparison with other works. We start with the work of S. Shinkevich and A. Studenikin~\cite{Sh05}. This work makes a similar analysis, but using a relativistic framework in free space. In free space, it is the total cross section the magnitude that makes sense. The spin asymmetry {\it A} (named as $S$ in that work), is taken as an input of the model ({\it i.e.} it is not explicitly evaluated). The spin asymmetry is incorporated to their results by making the replacement $s_n \to {\it A}$. In the absence of dense medium, this replacement leads to the correct expression. We have an overall agreement with their results, having in mind that in our case the effect of the dense medium is important and the comparison is only qualitative. In our case, a dense medium imposes restrictions on the available phase space, which depends on the temperature. The net effect is a smoothing of the results in relation to theirs. The work by D.A. Baiko and D.G. Yakovlev~\cite{Ba99}, a formalism similar to ours is employed. However, they focus on very low temperatures, being the scope of this paper different than ours. To the best of our knowledge perhaps the most complete analysis on the subject has been made by Maruyama~{\it et al.}~\cite{Ma12}. We should quote that we have obtained a general agreement with all these papers. What sets us apart from the other works is the treatment we make of the equation of state. We have determined the EoS with a magnetic field and from this we obtain spin--dependent single particle energies and a chemical potential which lead to specific values for $\rho_{+}$ and $\rho_{-}$, the density of neutron with spin up and down, respectively. Even though the spin asymmetry {\it A}, appears explicitly in the expression for the cross section, an accurate evaluation of the structure function requires single particles and chemical potential consistent with the value of the magnetic field.


\section{Summary and Conclusions}
\label{Summary}
In this work we have evaluated the neutrino mean free path for the absorption reaction $\nu  +  n  \to  e^{-} + p$, in hot dense neutron matter under a strong magnetic field. In first place, we have evaluated an EoS using the Hartree-Fock model with an Skyrme interaction with a strong magnetic field. As mentioned, we have a proton and an electron as final state. Being charged particles in a magnetic field, their quantum state is partially quantized, showing the so-called Landau levels. Due to this quantization, the phase space of final states is quite different from that of the same reaction but in the absence of a magnetic field. This contrast with the scattering reaction ($\nu  +  n  \to  \nu' + n'$), where the phase space of final states are very similar. While for B$=0$ the absorption reaction is always more important than the dispersion one, when B$\neq 0$, the situation is different: $\lambda_{abs}$ has a weak dependence with the temperature and decreases when the magnetic field grows, while $\lambda_{dis}$ has a strong dependence with the temperature (it decreases for growing values of T), and for $\theta_{\nu}=\pi/2$ is almost independent of the magnetic field. Therefore, in the presence of a strong magnetic field, either $\lambda_{abs}$ or $\lambda_{dis}$ can be the dominant contribution depending on the temperature. As a corollary of this behavior $\lambda_{abs}$ can be important for low temperatures as long as the magnetic field is strong.

For not null magnetic field, the neutrino mean free path depends on the angle between the neutrino momentum and the magnetic field (which we take as $\hat{z}$--axis). This establish a preferred direction in space resulting in an asymmetrical emission. This asymmetry is the result of the interplay among the weak, strong and electromagnetic interactions. The weak interaction is the responsible for the reaction $\nu  +  n  \to  \nu' + n'$, giving as a result a transition matrix element which depends on the spin of the particles involved. On the other hand, by solving the EoS for hot dense neutron matter under a strong magnetic field, we obtain a partially polarized system, from which we obtain single particle energies and the chemical potential needed for the evaluation of the neutrino mean free path. As already mentioned, the EoS gives us the equilibrium situation among the strong interaction (which favors ${\it A=0}$) and the coupling to the magnetic field (${\it A \to -1}$). It is worth to mention that this kind of analysis is quite involved for a more complex medium. If we simple add protons to the medium (see for instance~\cite{Ag15}), we need to work with two spin asymmetries: the one for neutrons and another one for protons. In this case, we already have Landau levels in the initial state and the whole scheme should be re--formulated.

Our results shows that the shortest neutrino mean free path is obtained for neutrinos moving anti--parallel to the magnetic field. As a consequence it is expected that the flux of emitted neutrinos parallel to the magnetic field is bigger than the one in the opposite direction. In Eq.~(\ref{chimfp}) we have defined the mean free path asymmetry $\chi_{tot}$, in order to account for this asymmetry in a quantitative way. We have obtained rather big values for $\chi_{tot}$. However, it would be speculative to draw a conclusion from these values: the geometry of the magnetic field in a neutron stars should be considered as well as the local density and temperature. Moreover, as discussed in the last paragraph, the actual composition of a neutron star is more complex. In any case, we consider that in the search for an explanation for the pulsar kick problem, this asymmetry can not be ignored.

In this work we have tried to give a self-consistent treatment of the mean free path for neutrinos, starting from the EoS and putting special emphasis in its asymmetry. Both the weak transition matrix element and the EoS contribute to the asymmetry in the neutrino mean free path. We have employed pure hot dense neutron matter due to it simplicity and because it is a reasonable assumption that this model represents one important contribution to the problem. Nuclear correlations beyond the mean field could have a relevant effect on the mean free path and its asymmetry. One way to deal with these correlations is the so-called ring approximation (see for instance~\cite{Pe09a,Pe09b}). But there are other correlations that can be also important. From this, our aim for a next work is to analyze the role of nuclear correlations beyond the mean field on the neutrino mean free path.

\newpage
\appendix
\section{Evaluation of the structure function $S_{s_p,s_n,N_p,N_e}$}
\label{strucfunc}
In this Appendix we evaluate the structure function for the absorption process $S_{s_p,s_n,N_p,N_e}$. We present a general expression, but at the end of this Appendix, we show a simpler expression which is more appropriate for our work. We recall the structure function defined in Eq.~(\ref{stfun0}),
\begin{eqnarray}
S_{s_p,s_n,N_p,N_e} & = & \int^{\infty}_{-\infty} \, \frac{d p_{\, n,z}}{2 \pi} \, \int^{\infty}_{-\infty} \, \frac{d p_{\, p,z}}{2 \pi} \,
(2 \pi)^{2} \, \delta(E_e+E_p-|p_{\nu}|-E_n) \nonumber \\
&\times& \delta(p_{\, e,z}+p_{\, p,z}-p_{\, \nu,z}-p_{\, n,z}) \; f_{s_n}(E_n,\mu_n,T) \, (1-f_{s_p}(E_p,\mu_p,T)),
\label{stfun1}
\end{eqnarray}
where $f_{s_i}(E_{i},\mu_i, T)$ has been given in Eq.~(\ref{fdd0}). The single-particle energies $E_{i}$ and the chemical potentials $\mu_i$ should be obtained from a particular model for the medium, which in our case is the Skyrme model (see \cite{Ag11,Ag14} and references therein). Within the Skyrme model, the nucleons single--particle energies for particles in a magnetic field, can be written as,
\begin{eqnarray}
E_{p}&=&m_p+\frac{p^{2}_{\, p,z}}{2 m^{*}_{s_p}}+\frac{e B}{m_p} \, (N_p+\frac{1}{2}) \, - \,s_p \mu_{Bp} B + \frac{ v_{s_p}}{8} \nonumber \\
E_{n}&=&m_n+\frac{p^{2}_{\, n}}{2 m^{*}_{s_n}}- \,s_n \mu_{Bn} B + \frac{ v_{s_n}}{8},
\label{spepn}
\end{eqnarray}
where $\mu_{Bp}$ and $\mu_{Bn}$ are the proton and neutron magnetic moments, respectively and $N_p$ indicates the Landau level. The effective masses ($m^{*}_{s_p}$ and $m^{*}_{s_n}$), together with the residual terms $v_{s_p}$ and $ v_{s_n}$, depend on the density of the system and explicit expressions are found in~\cite{Ag11,Ag14}. The structure function gives us information on the accessible phase--space of protons and neutrons. Even thought we work with neutron matter, the single particle energies in Eq.~(\ref{spepn}) are the ones for proton--neutron matter. We have employed these energies to give a more general expression for the structure function.

We take both the neutrino and the electron energies as in free space (with a magnetic field). We are considering massless neutrinos which are left-handed (or polarized). The energy of the electron is taken as,
\begin{equation}
E_{e}=(m^{2}_{e} + 2 e B N_e + p^{2}_{e, \, z})^{1/2}.
\label{spee}
\end{equation}
Note that due to the particular value for the magnetic moment of the electron, one can arrange the expression so that the energy depends only on $N_e$.

We now use the delta--function representing the momentum conservation in Eq.~(\ref{stfun1}), to obtain,
\begin{equation}
S_{s_p,s_n,N_p,N_e}  =  \int^{\infty}_{-\infty} \, d p_{\, n,z} \,
\delta(E_e+E_p-|p_{\nu}|-E_n) f_{s_n}(E_n,\mu_n,T) \, (1-f_{s_p}(E_p,\mu_p,T)),
\label{stfun2}
\end{equation}
where $p_{\, p,z}=p_{\, \nu,z}+p_{\, n,z}-p_{\, e,z}$. By
assigning impulse values to the lines in the diagram in Fig.~\ref{figme1}, energy--momentum conservation allow us to write,
\begin{eqnarray}
q_{0}&=& |p_{\nu}| - E_{e} \nonumber \\
q_{z}&=& p_{\, \nu,z}-p_{\, e,z} \nonumber \\
p_{\,p,z}&=&p_{\,n,z}+q_{z}.
\label{enemc}
\end{eqnarray}
Using these expressions we replace the energy and the $z$--momentum component of the electron by $q_{0}$ and $q_{z}$. The remainder integral in Eq.~(\ref{stfun2}) can be done, by solving the energy--conservation equation:
\begin{equation}
E_e+E_p-|p_{\nu}|-E_n = 0,
\label{enercon}
\end{equation}
which in fact, is a polynomial of second order in $p_{\, n,z}$. After some algebra, we have,
\begin{equation}
\alpha_{n} \, p^{2}_{\,n,z} \, + \, \beta_{n} \, p_{\,n,z} \,+ \, \gamma_{n} \, = 0,
\label{enercon2}
\end{equation}
where,
\begin{eqnarray}
\alpha_{n} &=& \frac{1}{2} \, (\frac{1}{m^{*}_{s_p}}-\frac{1}{m^{*}_{s_n}}) \nonumber \\
\beta_{n}  &=& \frac{q_z}{m^{*}_{s_p}} \nonumber \\
\gamma_{n} &=&  - \frac{p^{2}_{\, n,\perp}}{2 m^{*}_{s_n}} + \frac{q^{2}_{z}}{2 m^{*}_{s_p}}-m_n+m_p-q_0+\frac{e B}{m_p} \, (N_p+\frac{1}{2})
-s_p \mu_{Bp} B \nonumber \\
&+&s_n \mu_{Bn} B + \frac{1}{8} ( v_{s_p}- v_{s_n}),
\label{constesg}
\end{eqnarray}
We recall that $p_{\, n,\perp} = \sqrt{p^{2}_{\, n,x}+p^{2}_{\, n,y}}$. The energy--momentum of the neutrino and the electron enter into the structure function through the external quantities $q_0$ and $q_z$. This means that our expression for the structure function remains valid also for a dense system build up from protons, neutrons, electrons and neutrinos. Energy conservation can now be rewritten as,
\begin{equation}
\delta(E_p-E_n-q_0) = \frac{1}{(\beta^{2}_{n}-4 \alpha^{2}_{n} \gamma^{2}_{n})^{1/2}} \; [\delta(p_{\, n,z}-p^{+}_{\, n,z})+
\delta(p_{\, n,z}-p^{-}_{\, n,z})],
\label{deltaroot}
\end{equation}
where $p^{\pm}_{\, n,z}$ are the roots of Eq.~(\ref{enercon2}). Finally, the expression for the structure function is given by,
\begin{eqnarray}
S_{s_p,s_n,N_p,N_e} &  = & \frac{1}{(\beta^{2}_{n}-4 \alpha^{2}_{n} \gamma^{2}_{n})^{1/2}} \;
[f_{s_n}(E_n,\mu_n,T) \, (1-f_{s_p}(E_p,\mu_p,T))|_{p_{\, n,z}=p^{+}_{\, n,z}}  \nonumber \\
&+& f_{s_n}(E_n,\mu_n,T) \, (1-f_{s_p}(E_p,\mu_p,T))|_{p_{\, n,z}=p^{-}_{\, n,z}}].
\label{stfun3}
\end{eqnarray}
In particular, in this work we consider pure neutron matter. Therefore, in Eqs.~(\ref{spepn},\ref{constesg},\ref{stfun3}), we have to replace, $f_{s_p}(E_p,\mu_p,T)) \to 1$, $m^{*}_{s_p} \to m_{p}$ and $ v_{s_p} \to 0$, having,
\begin{equation}
S_{s_p,s_n,N_p,N_e}   =  \frac{1}{(\beta^{2}_{n}-4 \alpha^{2}_{n} \gamma^{2}_{n})^{1/2}} \;
[f_{s_n}(E_n,\mu_n,T)|_{p_{\, n,z}=p^{+}_{\, n,z}}+ f_{s_n}(E_n,\mu_n,T)|_{p_{\, n,z}=p^{-}_{\, n,z}}].
\label{stfun4}
\end{equation}
As mentioned in the text, we should recall that the structure function is a function of many variables. For simplicity, we show explicitly only the discrete variables, but it also depends on $q_0$, $q_z$, $p^{2}_{\, n,\perp}$, $m^{*}_{s_p}$, $m^{*}_{s_n}$, $\mu_p$, $\mu_p$, $T$ and $B$.

Another limit is when $m^{*}_{s_p}=m^{*}_{s_n}=m_N$. In this case, we have $\alpha_{n}=0$ and Eq.~(\ref{enercon2}) reduce to,
\begin{equation}
\beta_{n} \, p_{\,n,z} \,+ \, \gamma_{n} \, = 0,
\label{enercon4}
\end{equation}
that is, $p_{\,n,z}=-\gamma_{n}/\beta_{n}$ and
\begin{equation}
\frac{1}{(\beta^{2}_{n}-4 \alpha^{2}_{n} \gamma^{2}_{n})^{1/2}} \to \frac{m_N}{\mid q_z \mid}
\label{disc}
\end{equation}
and the structure function is,
\begin{equation}
S_{s_p,s_n,N_p,N_e}   =  \frac{m_N}{\mid q_z \mid} \;
f_{s_n}(E_n,\mu_n,T) \, (1-f_{s_p}(E_p,\mu_p,T))|_{p_{\, n, z}=-\gamma_{n}/\beta_{n}},
\label{stfun5}
\end{equation}
which is the same expression as in Eq.~(E2) in~\cite{Ar99}.

As a final comment on this Appendix, we should mention that for $\beta^{2}_{n}-4 \alpha^{2}_{n} \gamma^{2}_{n}=0$ (or equivalently for $q_z=0$ in Eq.~(\ref{stfun5})), there is a point for which the structure function is undefined. This is because at this point the energy has a double pole ($p^{+}_{\, n,z}=p^{-}_{\, n,z}$).

\section*{Acknowledgements}
We thank Isaac Vida\~na for his careful reading of our manuscript. This work was partially supported by the
CONICET, Argentina, under contract PIP00273.

\newpage

\newpage

\end{document}